\documentclass[aps,prb,twocolumn,groupedaddress,floatfix,showpacs]{revtex4}
\usepackage{latexsym}
\usepackage{dcolumn}
\usepackage[dvips]{graphicx}
\usepackage{amssymb}
\usepackage{graphics}
\usepackage{amsmath}
\usepackage{epsf}
\usepackage{bm}

\newcommand{\ty}[1]{\mbox{\tiny #1}}

\begin{document}

\author{Hongki Min$^{1,2}$}
\email{hmin@umd.edu}
\altaffiliation[Current address: ]
{Condensed Matter Theory Center, Department of Physics, 
University of Maryland, College Park, Maryland 20742, USA}
\author{Parakh Jain$^{1,3}$}
\author{S. Adam$^1$}
\author{M. D. Stiles$^{1}$}
\affiliation{$^1$ Center for Nanoscale Science and Technology, 
National Institute of Standards and Technology, 
Gaithersburg, Maryland 20899-6202, USA}
\affiliation{$^2$ Maryland NanoCenter, University of Maryland, 
College Park, Maryland 20742, USA}
\affiliation{$^3$Poolesville High School, 17501 West Willard Rd.
Poolesville, Maryland 20837, USA}

\title{Semiclassical Boltzmann transport theory for graphene multilayers}

\date{\today}
\begin{abstract}
We calculate the conductivity of arbitrarily stacked multilayer
graphene sheets within a relaxation time approximation, considering
both short-range and long-range impurities.  We theoretically investigate
the feasibility of identifying the stacking order of
these multilayers using transport measurements.  For relatively clean
samples, the conductivities of the various stacking configurations
depend on the carrier density as a 
power-law for over two decades.  This dependence
arises from a low density decomposition of the multilayer band
structure into a sum of chiral Hamiltonians.  For dirty samples, the
simple power-law relationship no longer holds.  Nonetheless,
identification of the number of layers and stacking sequence is still
possible by careful comparison of experimental data to the results
presented here.
\end{abstract}
\maketitle

\section{Introduction}
Enormous progress has been made in the five years since the first
experiments demonstrated that charge 
carriers in single graphene sheets behave like massless Dirac
fermions.  Making and studying 
such single-atom thick carbon sheets is now routinely done using a
wide variety of techniques.  (For reviews, see
Refs.~\onlinecite{kn:neto2009,kn:dassarma2010a}).

While the allure of manipulating single monoatomic sheets has
understandably attracted most of the attention in this field, from a
technological, or for that matter, fundamental point of view, the
properties of few-layer-graphene sheets are equally attractive.  Many
of the methods used to make monolayer graphene, such as mechanical
exfoliation of graphite, epitaxial growth from silicon carbide and
chemical vapor deposition on metals, can be suitably adapted to make
graphene stacks, with a controllable number of atomic layers.  Many of
the unusual properties of the Dirac Hamiltonian that are used to
describe monolayer graphene, such as having chiral and ambipolar
carriers, survive in these multilayers, so one might expect that these
sheets would have high mobility and favorable carrier transport
properties.  
As a result, multilayer graphene may play an important role in future
electronic devices where its additional ``layer'' degree of freedom
could be manipulated~\cite{kn:avetisyan2010} to achieve desirable properties, such as the 
demonstrated gate-tunable band-gap in graphene 
bilayers.~\cite{kn:zhang2009b} 

There has been experimental and theoretical work on
the optical properties of graphene multilayers, as well as some very recent theoretical
predictions on the phonon-scattering in these
multilayers.\cite{kn:min2010} However, there has not been a systematic
study of the low temperature transport properties of graphene
multilayers.  In
anticipation of forthcoming experiments on these systems, we use both
analytical and numerical methods to understand carrier transport in
graphene multilayers.

The complication in studying multilayers is the coupling between 
the layers.  The carrier transport in a single graphene sheet can be readily
understood using the Dirac Hamiltonian, which is the low energy
effective theory for $\pi$-orbitals located on the vertices of a
carbon honeycomb lattice.  For multilayers, the additional coupling 
between orbitals on neighboring layers depends sensitively 
on many factors such as the distance
between the layers and their relative orientation.  For
example, graphene bilayers with a twist angle between their respective
primitive cells are predicted to largely act as decoupled
sheets.\cite{kn:hass2008,kn:santos2007}  For Bernal stacking
(also called A-B stacking), on the
other hand, half the carbon atoms in each
hexagon of the top layer lie exactly over the center of a hexagon of
the layer below it.  The resulting strong coupling between the 
two layers gives a low energy effective theory with a  
zero-gap hyperbolic dispersion.  

While height fluctuations (or equivalently,
having spatial fluctuations in the interlayer coupling strength), or
allowing for arbitrary rotations and slips between the layers are
important for some systems (such as epitaxial graphene), their effects
are beyond the scope of the present
work.\cite{kn:mele2010,kn:bistritzer2010}  Here we consider multilayers 
that come in families where the orientation of the upper layers
is determined by symmetry considerations from the orientation of the
bottom layer. This would be the case, for example, if the multilayer
inherits its structure from a parent structure, as is the case in the
mechanical exfoliation technique.

We further restrict our multilayer analysis to the lower energy
stacking sequences in which neighboring layers share only one sublattice.
For example, we consider bilayer graphene that is A-B stacked, where 
the two layers share one sublattice, but not A-A stacked,
where each carbon atom of the top layer lies exactly on top of a
carbon atom of the bottom layer (sharing both sublattices).  The 
consecutive A-A stacking is 
energetically unfavorable,\cite{kn:charlier1992} so we do 
not consider this stacking and its generalization in multilayer stacks.
For trilayer graphene, we consider two possibilities: 
A-B-A stacking (also called Bernal-like) is a Bernal bilayer, with the 
third layer having carbon atoms located directly above the bottom layer; 
and A-B-C stacking.

\begin{figure}[t]
\begin{center}
\includegraphics[width=0.7\linewidth]{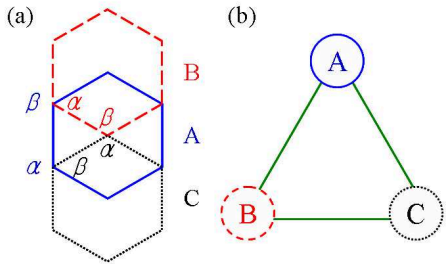}
\caption{(Color online) (a) Schematic illustration of (a) three types
  of stacking arrangements, labeled by A, B and C. The honeycomb
  lattice of a single sheet has two triangular sublattices, labeled by
  $\alpha$ and $\beta$. (b) Each added layer cycles around this
  stacking triangle in either the right-handed or the left-handed
  sense.  Reversals of the sense of this rotation tend to increase the
  number of low-energy pseudospin doublets.}
\label{fig:stack}
\end{center}
\end{figure}

Figure \ref{fig:stack} illustrates the different stacking sequences 
graphically.  Since the honeycomb lattice of a single graphene sheet comprises
two interpenetrating triangular sublattices, we label the 
sublattices of each layer $\alpha$ and $\beta$.  When a subsequent graphene 
layer is placed on top of the stack, we consider the stacking orders where either
the atoms of the $\alpha$ or the $\beta$ sublattices are displaced 
along the edges of the honeycomb of this top sheet.  This gives a 
stacking rule that implies three distinct but equivalent projections 
(labeled A, B, and C) of the three-dimensional structure's honeycomb-lattice 
layers onto the $\hat{x}$-$\hat{y}$ plane and consequently $2^{N-2}$ distinct 
stacking sequences for an $N$-layer stack.  

The electronic properties of multilayer graphene strongly depend on the
stacking sequence.  Periodically stacked multilayer graphene
\cite{kn:latil2006,kn:guinea2006} and arbitrarily stacked multilayer
graphene \cite{kn:min2008} have been studied theoretically,
demonstrating that the low-energy band structure of a graphene
multilayer consists of a set of independent pseudospin doublets. It
was shown that an energy gap can be induced by a perpendicular external
electric field in ABC-stacked multilayer 
graphene.\cite{kn:aoki2007,kn:zhang2010} 
Furthermore, in ABC stacking, electron-electron interactions play a more important 
role than other stacking sequences due to the appearance of relatively flat bands near the Fermi level.\cite{kn:zhang2010}  
This enhanced role of electron interactions raises the likelihood of strongly-correlated ground-states, 
a possibility that we ignore in our semiclassical treatment below.
Optical properties of multilayer
graphene using absorption spectroscopy have been studied
experimentally \cite{kn:mak2010} and theoretically
\cite{kn:koshino2009,kn:min2009} showing characteristic peak positions
in optical conductivity depending on the stacking sequence. 

Transport properties of monolayer, bilayer and multilayer graphene
have been studied
theoretically\cite{kn:nilsson2006,kn:nilsson2007,kn:nilsson2008,
  kn:yuan2010} within coherent potential approximations.  These
approaches capture the scattering properties of the impurity
potential (which is important for strong disorder), but they are often
restricted to small system sizes, and do not accurately account for
the disorder-induced spatial inhomogeneity of the fluctuating local
carrier density. We believe this inhomogeneity dominates the transport
properties of these graphene multilayers (see discussion in
Ref.~\onlinecite{kn:dassarma2010a}).  

Our main finding is that for relatively clean samples, 
the carrier density dependence of the multilayer conductivity follows
a power-law dependence for more than two decades, a direct
consequence of the effective low energy chiral decomposition.  For dirty samples,
the carrier density inhomogeneity induced by the disorder washes away
this power-law relationship.  However, the various stacking sequences
give characteristically different dependence of the multilayer
conductivity on carrier density.  By careful comparison with
experimental data, our results could be used to identify both the
number of layers and the stacking sequence of a multilayer graphene
sample.
   
The rest of this manuscript is organized as follows.  In
Sec.~\ref{model}, we describe the theoretical model where we solve for
the multilayer graphene band structure using a tight-binding model
that includes both the nearest-neighbor intralayer hopping and the
nearest-neighbor interlayer hopping, and solving for the conductivity
within the Boltzmann transport formalism.  In Sec.~\ref{results} we
present our results for graphene stacks comprising one, two, three and
four layers, treating impurity scattering by both Coulomb potentials
and short-range disorder. In the 
appendices we present details of the chiral decomposition and the
transport properties of $J$-chiral fermions, as well as analytic results
for the electronic transport in bilayer graphene.

\section{Theoretical model}       
\label{model}

\subsection{Tight binding Hamiltonian}
\label{tightbinding}

The low energy effective Hamiltonian for the $\pi$-orbital continuum
model for arbitrarily stacked $N$-layer graphene centered at the
hexagonal corners of the Brillouin zone is given by~\cite{kn:partoens2006,kn:partoens2007}
\begin{equation}
\label{eq:Hamiltonian}
{\cal H}=\sum_{\bm p} \Psi_{\bm p}^{\dagger} H({\bm p}) \Psi_{\bm p},
\end{equation}
where $\Psi_{\bm p}=(c_{1,\alpha,{\bm p}},c_{1,\beta,{\bm p}},\cdots,c_{N,\alpha,{\bm p}},c_{N,\beta,{\bm p}})$ and $c_{l,\mu,{\bm p}}$ is an electron annihilation 
operator for layer $l=1,\cdots,N$, sublattice $\mu=\alpha,\beta$ and 
momentum $\bm p$ measured from $K$ or $K'$ point.

The simplest model for a multilayer graphene system allows only
nearest-neighbor intralayer hopping $t$ and the nearest-neighbor
interlayer hopping $t_{\perp}$. The in-plane Fermi velocity for
monolayer graphene, $v_0$, is
related to $t$ by ${\hbar v_0 \over a}={\sqrt{3}\over 2} t$, where $a=0.246$
nm is the lattice constant of monolayer graphene.  This model ignores
some aspects of the electronic band structure -- in principle,
corrections to this model, such as adding next-nearest-neighbor
hopping, can easily be included, although in practice, it is often
numerically quite intensive.  We find that such corrections do not
significantly alter any of our main findings.

\subsection{Boltzmann transport theory}
The conductivity is a property of electrons close to the 
Fermi energy and is obtained from the Einstein relation,
$\sigma_{\rm B}=e^2 {\cal D}(\varepsilon_{\rm F}) D$ where ${\cal D}(\varepsilon_{\rm F})$ 
is the density of states at the Fermi energy $\varepsilon_{\rm F}$ and $D$ is 
the diffusion constant.  
In cases in which the Fermi surface has multiple sheets (lines in the two dimensional cases we consider here), 
the conductivity is the sum over such contributions for each sheet.  See Sec.~\ref{intervalley_interband} for details.
For convenience, we denote the density of states for each sheet as
${\cal D}(\varepsilon_{\rm F})=g_{\rm s} g_{\rm v} \rho(\varepsilon_{\rm F})$
where $g_{\rm s}=2$ and $g_{\rm v}=2$ are spin and valley degeneracy factors.
For diffusive transport in two dimensions, $D={1\over 2} v_{\rm F}^2 \tau_{\rm F}$.
The Fermi velocity, density of states and 
relaxation time can be calculated from the dispersion relation as
\begin{subequations}
\begin{equation}
v_{\rm F}={1\over \hbar}\left.{d\varepsilon\over
  dk}\right|_{\varepsilon=\varepsilon_{\rm F}},
\label{eq:vF}
\end{equation}
\begin{equation}
\rho(\varepsilon_{\rm F})={k_{\rm F} \over 2\pi\left|{d\varepsilon/
    dk}\right|_{\varepsilon=\varepsilon_{\rm F}} }={k_{\rm F} \over 2\pi \hbar   v_{\rm F}},
\label{eq:rho}
\end{equation}
\begin{equation}
{1\over \tau_F}={2\pi \over \hbar} n_{\rm imp} V_{\rm imp}^2 \rho(\varepsilon_{\rm F}),
\label{eq:tau}
\end{equation}
where $V_{\rm imp}^2$ is the squared effective impurity scattering potential 
averaged over the angle, and the Fermi wavevector is related to the carrier density 
and applied back gate voltage $k_{\rm F}^2 = 4\pi n/(g_{\rm s} g_{\rm v}) \propto V_G$.  
When the Fermi surface has multiple sheets, the left hand side of this
last relation should be summed over $k_{\rm F}^2$ for each sheet.
The validity of the Born approximation implicit in this formulation is discussed in Appendix~\ref{Ap:born}, 
and the special case of $J$-chiral fermions is treated in Appendix~\ref{Ap:chirality}.
\end{subequations}

The scattering matrix element $V_{\rm imp}^2$ that gives rise to the transport relaxation 
time is obtained using the Boltzmann transport formalism
\begin{equation}
\label{eq:squared_potential}
V_{\rm imp}^2=\int_{0}^{2\pi} {d\phi \over 2\pi} |V_{\rm imp}(\phi)|^2
F(\phi)(1-\cos\phi),
\end{equation}
where $V_{\rm imp}(\phi)$ is the matrix element of the impurity potential at
scattering angle $\phi$, and $F(\phi)$ is the chirality factor that
arises from the projection of the spinor wavefunctions between the incoming
and outgoing states. For the case of intervalley and interband scattering, the
treatment of the chirality factor is more subtle and discussed in 
Appendix~\ref{Ap:chiralfactor}.  For convenience, we define the dimensionless 
potential $\tilde{V}_{\rm imp}(\phi)$
\begin{equation}
\label{eq:unitless_potential}
V_{\rm imp}(\phi)={2\pi e^2 \over \epsilon k_{\rm F}} \tilde{V}_{\rm imp}(\phi)={2\pi \hbar
  v \alpha \over k_{\rm F}} \tilde{V}_{\rm imp}(\phi),
\end{equation}
where $\epsilon$ is the dielectric constant and $\alpha={e^2 \over
  \epsilon \hbar v_0}$ is the effective fine structure constant.  The conductivity can then 
be written as   
\begin{equation}
\label{eq:conductivity}
\sigma_{\rm B} (n) ={e^2\over h}\left({n\over n_{\rm imp}}\right)\left({v_{\rm F}\over v_0}\right)^2 \left({1 \over 2\pi\alpha^2 \tilde{V}_{\rm imp}^2}\right),
\end{equation}
where $n=g_{\rm s} g_{\rm v} k_{\rm F}^2/(4\pi)$ is the carrier density, and all the information about the 
band structure and the nature of
the disorder potential is captured by $v_{\rm F}$ and ${\tilde V}_{\rm imp}^2$.
When several bands cross the Fermi energy, we calculate $v_{\rm F}$ and
${\tilde V}_{\rm imp}^2$ (which yields $\sigma_i$) for each of the $i$ bands and then 
calculate the total conductivity as $\sigma=\sum_i
\sigma_i$, where the applied gate voltage is proportional to $n
=\sum_i n_i$, and $n_i = g_{\rm s} g_{\rm v} k_{{\rm F}, i}^2/(4\pi)$ is fixed by keeping
$\varepsilon_{\rm F}$ the same for all bands.    

\subsection{Impurity Scattering}

Different types of impurity potentials give qualitatively different
results for the conductivity.  This can be seen in
Eq.~\ref{eq:conductivity}, where the wavevector dependence of the Fourier
transform of the impurity potential $V_{\rm imp}[q= 2 k_{\rm F} \sin(\phi/2)]$
changes the scaling of the conductivity $\sigma(n)$.
Studies on monolayer graphene have explored a wide variety of disorder
potentials including long-range Coulomb ($V_{\rm imp}(q) \sim q^{-1}$),
Gaussian-white noise ($V_{\rm imp}(q) \sim q^{0}$) and Gaussian-correlated
disorder ($V_{\rm imp}(q) \sim \exp[-q^2]$) as well as resonant scatterers that cause
a maximal phase shift of $\pi/2$ between incoming and outgoing
wavefunctions.  We refer to Ref.~\onlinecite{kn:dassarma2010a} for
more details on the different kinds of disorder in graphene.

For our purposes, we focus on what we believe to be the most relevant
scattering mechanisms, i.e. charged impurities (which act as screened
long-range Coulomb disorder) and short-range defects (approximated as
white-noise disorder).  Using the Thomas-Fermi screening theory, one 
can write expressions for the dimensionless scattering potential 
$\tilde{V}_{\rm imp}^2$ which was defined in Eq.~\ref{eq:unitless_potential}.  
For screened charged impurities from $V_{\rm imp}(\phi)={2\pi e^2 \over \epsilon q}e^{-qd_{\rm imp}}$ with $q=2 k_{\rm F} \sin(\phi/2)$ we find
\begin{equation}
\label{eq:squared_potential_screen}
\tilde{V}_{\rm imp}^2=  \int_{0}^{2\pi} {d\phi \over 2\pi} 
{F(\phi)(1-\cos\phi) \over \left(2 \sin(\phi/2)+\tilde{q}_{\rm TF}\right)^2 } e^{-4k_{\rm F} d_{\rm imp}\sin(\phi/2) },
\end{equation}
where the dimensionless Thomas-Fermi wavevector is 
$\tilde{q}_{\rm TF}= q_{\rm TF}/k_{\rm F} = g_{\rm s} g_{\rm v} \alpha (v_0/v_{\rm F})$
and $d_{\rm imp}$ is the distance between the impurities and the graphene sheet.
When several bands cross the Fermi energy, the Thomas-Fermi screening wavevector determined
from the \emph{total} density of states including all the bands.
The monolayer Fermi velocity $v_0$ was defined below Eq.~\ref{eq:Hamiltonian}.
  
The short-range disorder potential can be characterized by an 
effective scattering cross-section length $d_{\rm sc}$ such that
$V_{\rm imp}(\phi)={2\pi e^2 d_{\rm sc} \over \epsilon}$ or
\begin{equation}
\label{eq:squared_potential_short}
\tilde{V}_{\rm imp}^2=k_{\rm F}^2 d_{\rm sc}^2 \int_{0}^{2\pi} {d\phi \over 2\pi}  F(\phi)(1-\cos\phi).
\end{equation}
\noindent  Taken together with Eq.~\ref{eq:conductivity}, this completely 
defines the electrical conductivity in terms of the multilayer band structure 
discussed in Sec.~\ref{tightbinding}.

\subsection{Intervalley and Interband Contributions}
\label{intervalley_interband}

At typical carrier densities, the band structure of monolayer graphene
comprises two Dirac cones that are centered at two inequivalent points
(also called valleys) of the Brillouin zone boundary labeled as $K$
and $K'$.  The scattering of carriers between the valleys requires a
momentum transfer of $Q = 4 \pi /(3 a) \approx 17~{\rm nm}^{-1}$, 
and is strongly suppressed for Coulomb impurities.  For
short-range scattering, the treatment depends on how one models the
intervalley scattering matrix element, but in most cases it is
sufficient to consider a single valley with a suitably defined
impurity concentration $n_{\rm imp}$ (see e.g. Ref.~\onlinecite{kn:shon1998}).
For concreteness, we assume that for short-range impurities, the
intervalley and intravalley scattering matrix elements are equal,
that $n_{\rm imp}$ is the average impurity concentration in a single
valley (and is the same for both valleys).  The conductivity using
these definitions is smaller by a factor of 2 from the case of $n_{\rm
  imp} = n_{\rm tot} = n_A + n_B$, where $n_A$ and $n_B$ are
concentration of impurities or defects on the A and B sublattices,
respectively.

As discussed in Appendix~\ref{Ap:chirality}, within a single valley,
all finite stackings that are a subsequence of periodic ABC stacking; i.e. 
A, AB, ABC, ABCA, etc.; have only a
single band at low energies, and the chirality increases as the number of layers
increases.  However, for other stackings, the band
structure features multiple chiral bands with different dispersion
relations that are centered at the Dirac point.  For Coulomb
impurities, the matrix element in Eq.~\ref{eq:squared_potential_screen} can be computed for both
intraband and interband scattering, and their scattering rates
added in accordance with Matthiessen's rule.  For short-range
impurities, the interband scattering is more subtle.  A
straightforward application of Eq.~\ref{eq:squared_potential_short} would lead to a strong
suppression of all interband scattering because
the chirality factor $F(\phi)$ vanishes or is significantly less than
one.  As discussed in Appendix~\ref{Ap:chiralfactor} we 
believe this to be unphysical because it requires the short-range impurity
potential to be diagonal in the space of all the layer and
valley quantum numbers.  Most short range scatters we can imagine
would be localized to one layer and a specific sublattice so the
scattering potential would {\em not} be diagonal in this space.
The screened Coulomb potential, on the other hand, varies slowly
between the layers and sublattices and can be much better approximated
as diagonal in that space. 
In the absence of a microscopic model for a
particular impurity model (e.g. computing the scattering potential
resulting from a single vacancy or from the binding of a single
hydrogen atom to the top layer), we believe that for short range
scatterers, a more realistic
assumption is to set $F(\phi)=1$.  This correctly weights the relative
importance of intraband and interband scattering, and therefore
gives the correct qualitative carrier density dependence of the
conductivity.     

The role of interband scattering is most striking when one considers 
the large density regime where higher energy bands become accessible.  
For screened charged impurities, the additional density of states in these 
higher energy bands {\it enhance} the screening 
of long-range impurities which will sharply {\it increase} the conductivity,
while the interband scattering is suppressed by the chirality factor.  On the 
other hand, for short-range impurities, the higher energy band becomes an additional
 source of interband scattering that sharply {\it decreases} the conductivity.  At the
time of writing, there have been no transport experiments that could probe 
these higher energy bands by inducing sufficiently large carrier densities.  However, if
such an experiment is done in the future (perhaps by finding better electrolytes), then 
the increase (decrease) of $\sigma(n)$ would be indicative of Coulomb (short-range) 
impurities being the dominant source of scattering.   

\subsection{Effective Medium Theory (EMT)}

At low carrier density, the disorder induced fluctuations in the local
density become larger than the spatially averaged carrier density.
This has been called the electron-hole puddle regime.  We use the
effective medium approach developed in Ref.~\onlinecite{kn:rossi2008b}
to obtain the bulk conductivity $\sigma_{\ty{EMT}}$ of this
inhomogeneous medium.  It was shown in Ref.~\onlinecite{kn:adam2009b} 
that by assuming a Gaussian probability distribution for the carrier
density, $\sigma_{\ty{EMT}}(n)$ could be obtained from $\sigma_{\rm B}(n)$ using
\begin{equation}
\label{eq:emt1}
\int_0^\infty dn' \exp\left[\frac{-n'^2}{2 n_{\rm rms}^2}\right]
\cosh\left[\frac{n n'}{n_{\rm rms}^2}\right] \frac{\sigma_{\rm B}(n')
  - \sigma_{\ty{EMT}}(n)}{\sigma_{\rm B}(n') + \sigma_{\ty{EMT}}(n)} =
0, 
\end{equation}
 where $n_{\rm rms}$ parameterizes the Gaussian distribution, and
 $\sigma_{\rm B}(n)$ is obtained numerically from Eq.~\ref{eq:conductivity}.
 This integral equation extrapolates from $\sigma_{\ty{EMT}}(n) \approx
 \sigma_{\rm min}$ for $|n| \lesssim n_{\rm rms}$ to $\sigma_{\ty{EMT}}(n)
 \approx \sigma_{\rm B}(n)$ for $|n| \gg n_{\rm rms}$.  The effect of 
the puddles, therefore, is to give rise to a minimum conductivity plateau 
where the conductivity remains roughly constant when the average 
of the carrier density is smaller than its fluctuations.\cite{kn:adam2007a}

\section{Results and Discussion}
\label{results}

As outlined above, for an arbitrary graphene stack we first solve
Eq.~\ref{eq:Hamiltonian} to obtain the band structure.  For
simplicity, we choose $t=3$~eV, $t_\perp=0.3$~eV, $\alpha = 1$ and 
$n_{\rm rms} = n_{\rm imp}$.  Neglecting higher order hopping terms keeps the spectrum
rotationally symmetric (reducing the computational time), while the
ratio of $n_{\rm rms}/n_{\rm imp}$ is a number of order unity that can
be calculated within the low-density chiral decomposition (see
Appendix~\ref{Ap:chirality}).  We do not believe that these
approximations significantly alter our findings.

Having solved numerically for the wavefunctions, one can compute
$v_{\rm F}$ (Eq.~\ref{eq:vF}), $\rho(\varepsilon)$ (Eq.~\ref{eq:rho}),
as well as the chirality factor $F(\phi)$ (see Appendix~\ref{Ap:chiralfactor}).  
For the short-range and Coulomb disorder
potential we use representative~\cite{kn:jang2008} 
parameters: $d_{\rm sc}=0.3$ nm and $d_{\rm imp}=1$ nm.
Equation~\ref{eq:emt1} then gives the predicted carrier density
dependence of the conductivity. 

\begin{figure*}[!ht]
\bigskip
\begin{center}
\begin{tabular}{cc}
\includegraphics[width=0.45\linewidth]{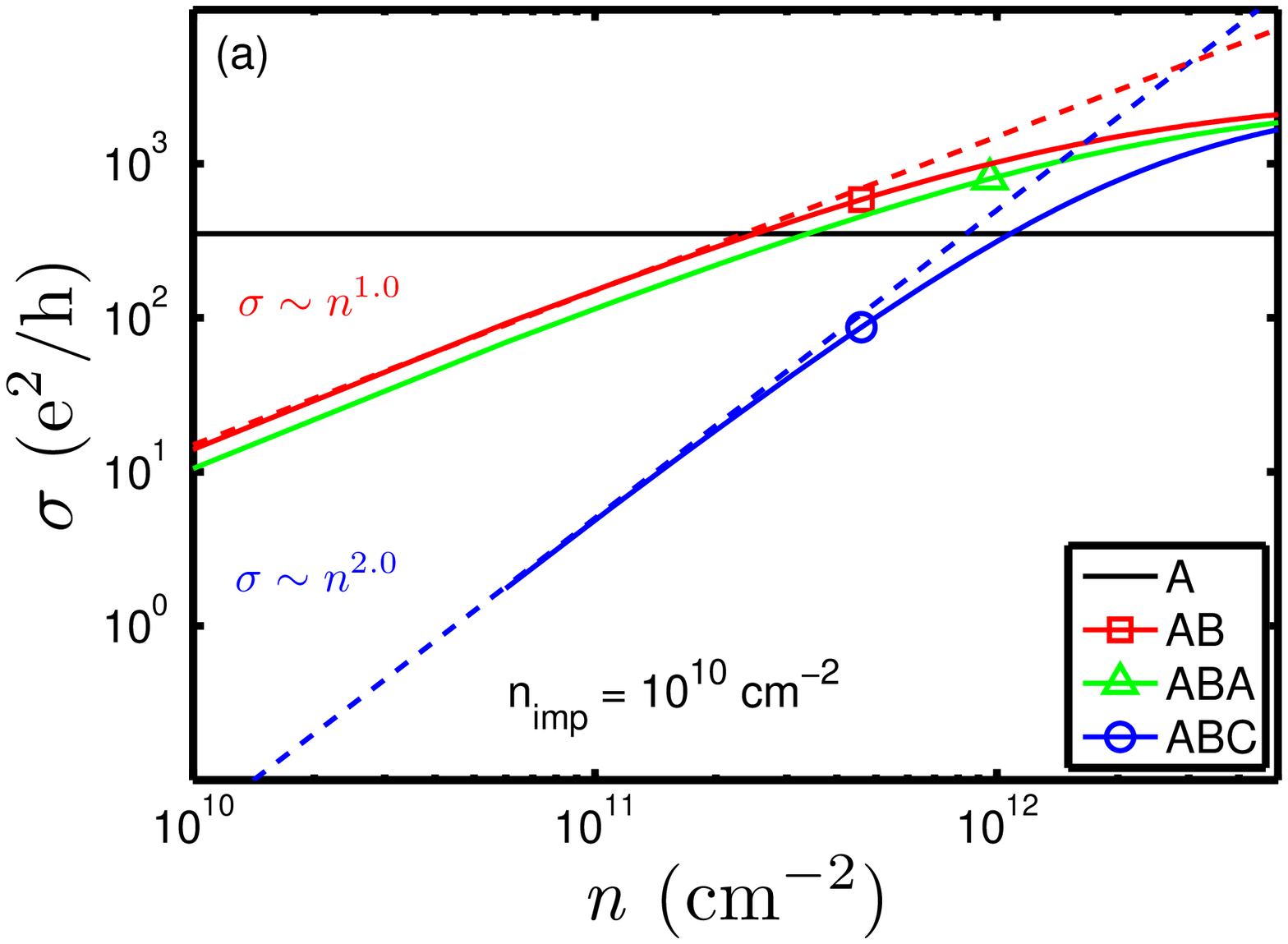} &
\includegraphics[width=0.45\linewidth]{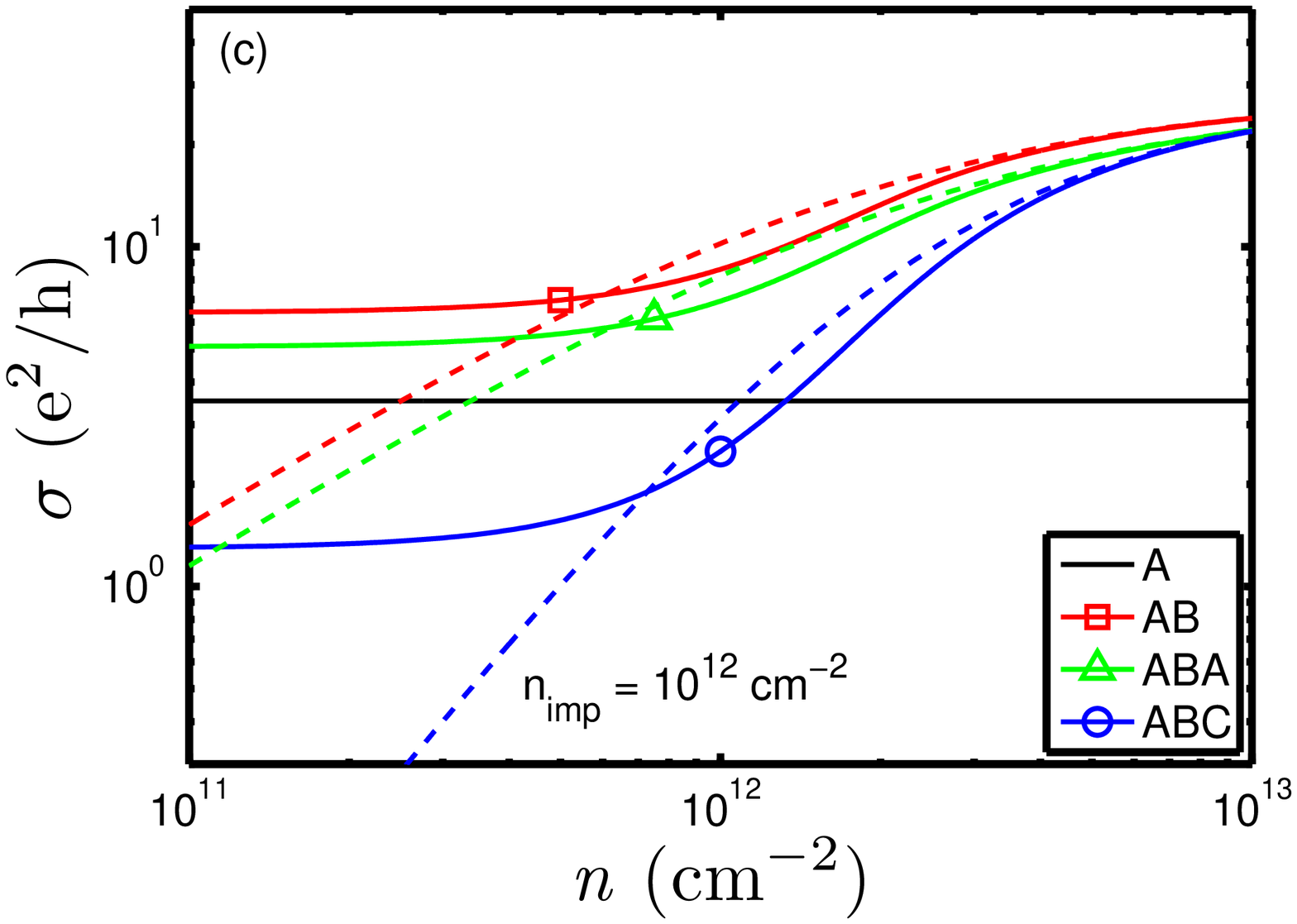} \\
\includegraphics[width=0.45\linewidth]{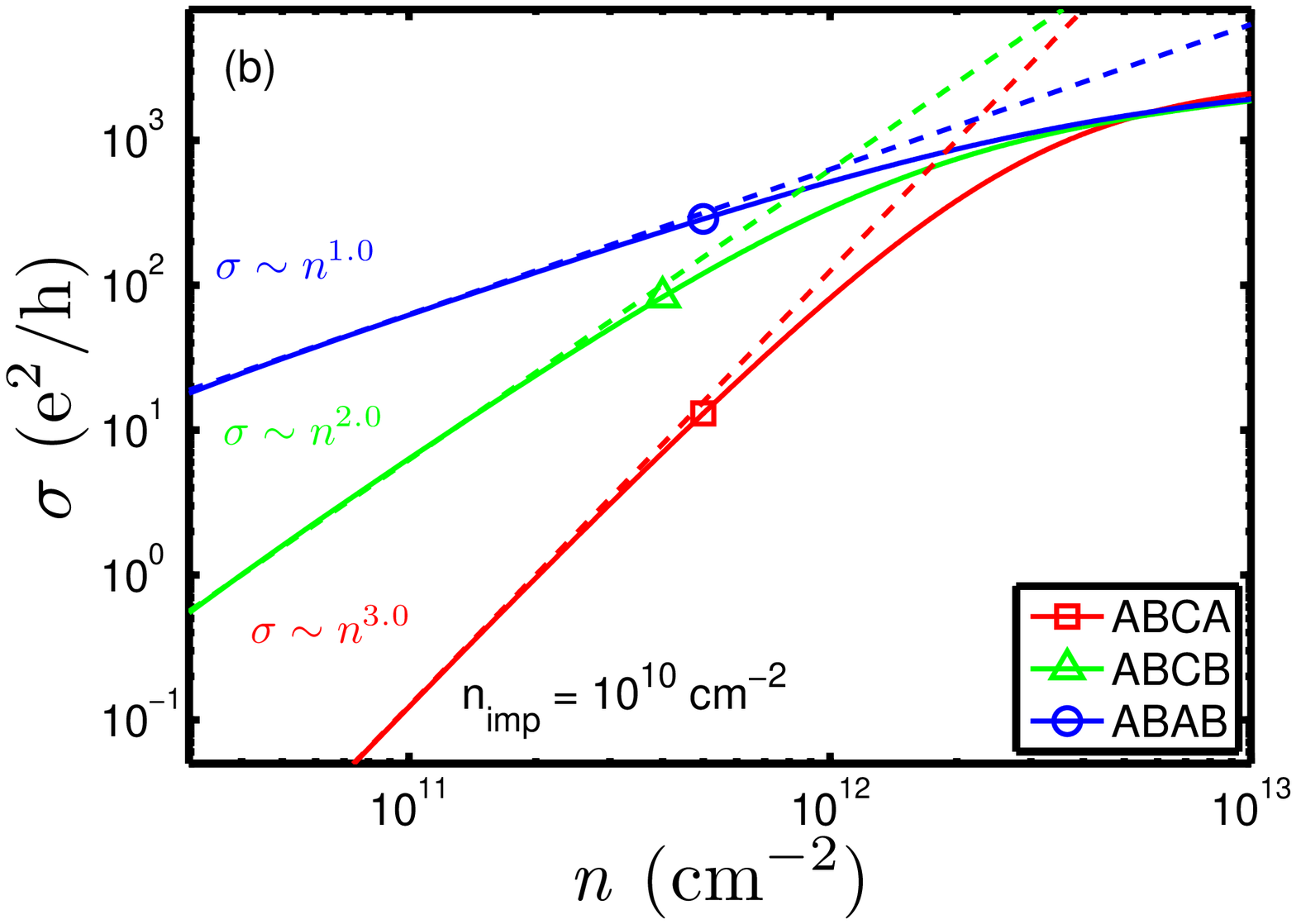} &
\includegraphics[width=0.45\linewidth]{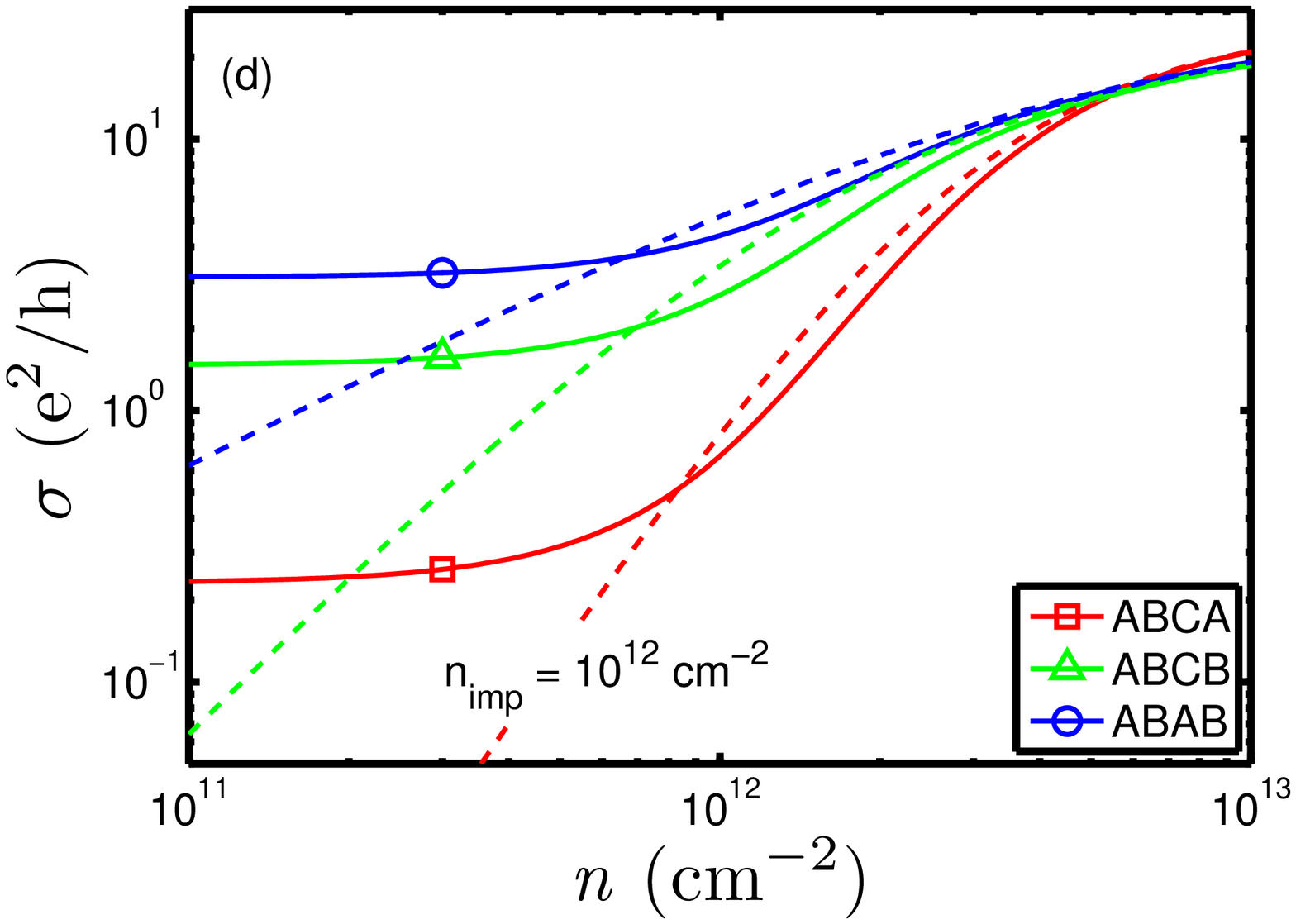}
\end{tabular}
\caption{\label{fig:short-range} (Color online) 
Graphene conductivity assuming short-range scatterers.  For clean
samples, ($n_{\rm imp} \lesssim 10^{10}~{\rm cm}^{-2}$), the
conductivity $\sigma_{\ty{EMT}}(n)$ has about two decades of 
power-law dependence.  (a) The conductivity (solid lines) for monolayer,
bilayer and ABC trilayer graphene each follow different power-laws (dashed lines). 
(b) The conductivity (solid lines) for the different stacking 
sequences for tetralayer
graphene also have different power-laws (dashed lines).  This
indicates that for short-range impurities, in most cases transport measurements can
be used to identify the number of graphene layers.  For dirty samples
(c) and (d), where $n_{\rm imp}= 10^{12}~{\rm cm}^{-2}$, the density
dependence is no longer given by a power-law.  Solid lines show the
conductivity after using an effective medium theory to average over
the disorder induced carrier density fluctuations.  Dashed lines are
the results before such averaging.  Although the conductivity does not
have power-law dependence on carrier density, the transport properties
still strongly depend on the number of layers and their relative
stacking-order.  Therefore, transport measurements could still be used 
to identify the type of graphene multilayer.}
\end{center}
\end{figure*}

Figure~\ref{fig:short-range} shows the results for short-range
scatterers such as point defects.  The left panel
(Fig.~\ref{fig:short-range}a and Fig.~\ref{fig:short-range}b) assumes
a relatively clean sample with $n_{\rm imp} = 10^{10}~{\rm cm}^{-2}$.
The solid lines are $\sigma_{\ty{EMT}}$ calculated from
Eq.~\ref{eq:emt1}, while dashed lines show the (approximate) power-law
dependence of the conductivity on carrier density.  We find that for
most cases the conductivity limited by short-range scatterers exhibits
a unique power-law (for more than two decades) that depends on the
number of layers and the stacking sequence.  For such clean samples,
where the scattering is dominated by short-range disorder, with the
exception of the similarity between AB bilayers and ABA trilayers, the distinct 
power-law dependence for $\sigma(n)$ enables easy identification of the sample
from transport measurements.

The right panel (Fig.~\ref{fig:short-range}c and Fig.~\ref{fig:short-range}d)
shows the results for
a dirty sample ($n_{\rm imp} = 10^{12}~{\rm cm}^{-2}$).  The solid lines
show the EMT result (Eq.~\ref{eq:emt1}) and the dashed lines show 
the Boltzmann result before the EMT averaging (Eq.~\ref{eq:conductivity}).
We note that the results for monolayer and bilayer graphene agree with 
analytical calculations discussed in Appendix~\ref{Ap:bilayer}.  While there 
is no simple power-law dependence of the conductivity on carrier density, the 
different stacking sequences have very different functional forms
for $\sigma(n)$.  (For example, the ratio $\sigma(n=10^{13}~{\rm cm}^{-2})/
\sigma(n=10^{11}~{\rm cm}^{-2})$ varies by almost an order of magnitude).
It might therefore still be possible to identify the number of 
layers and the stacking sequence from transport.               

\begin{figure*}[!ht]
\bigskip
\begin{center}
\begin{tabular}{cc}
\includegraphics[width=0.45\linewidth]{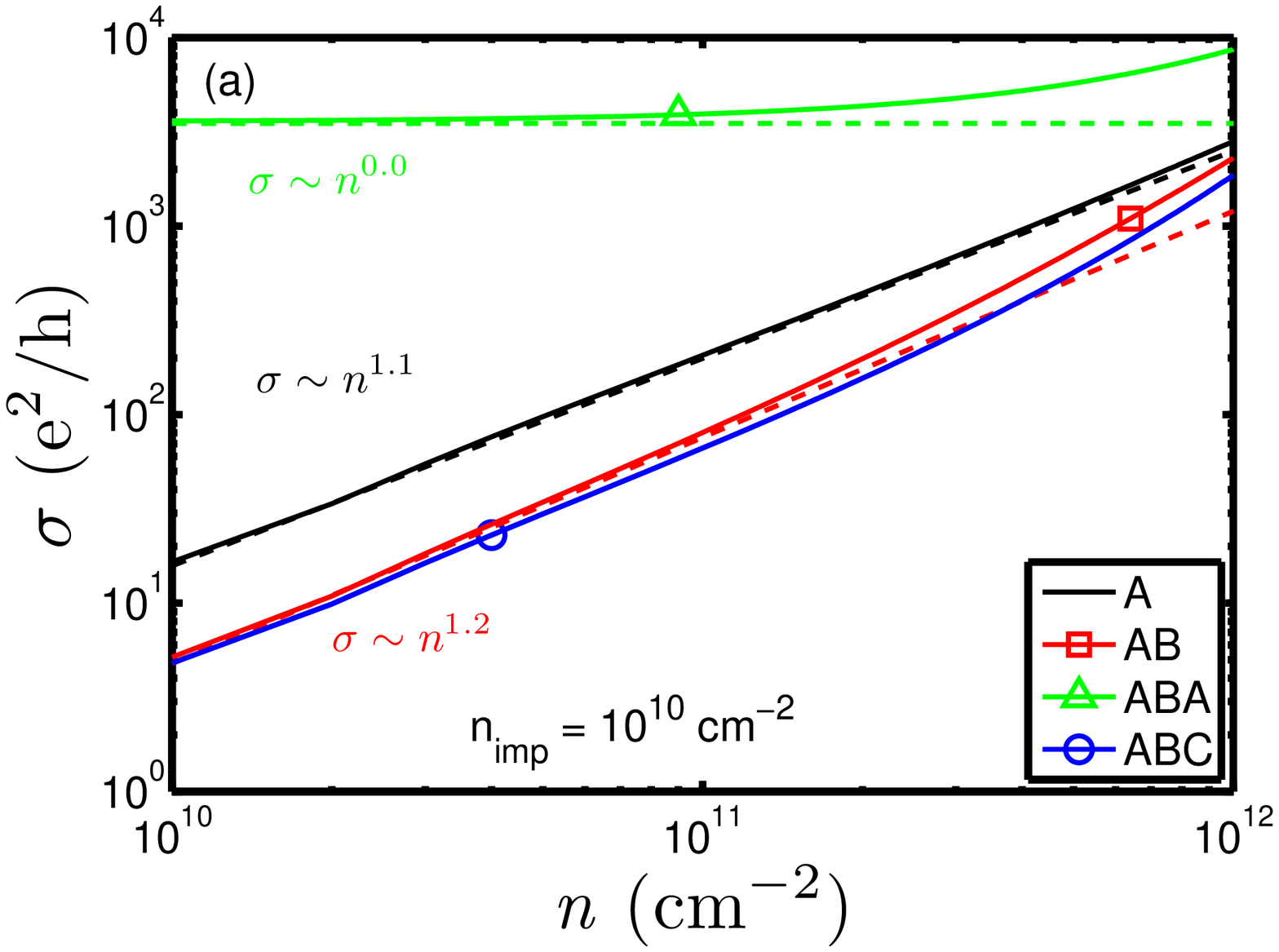} &
\includegraphics[width=0.45\linewidth]{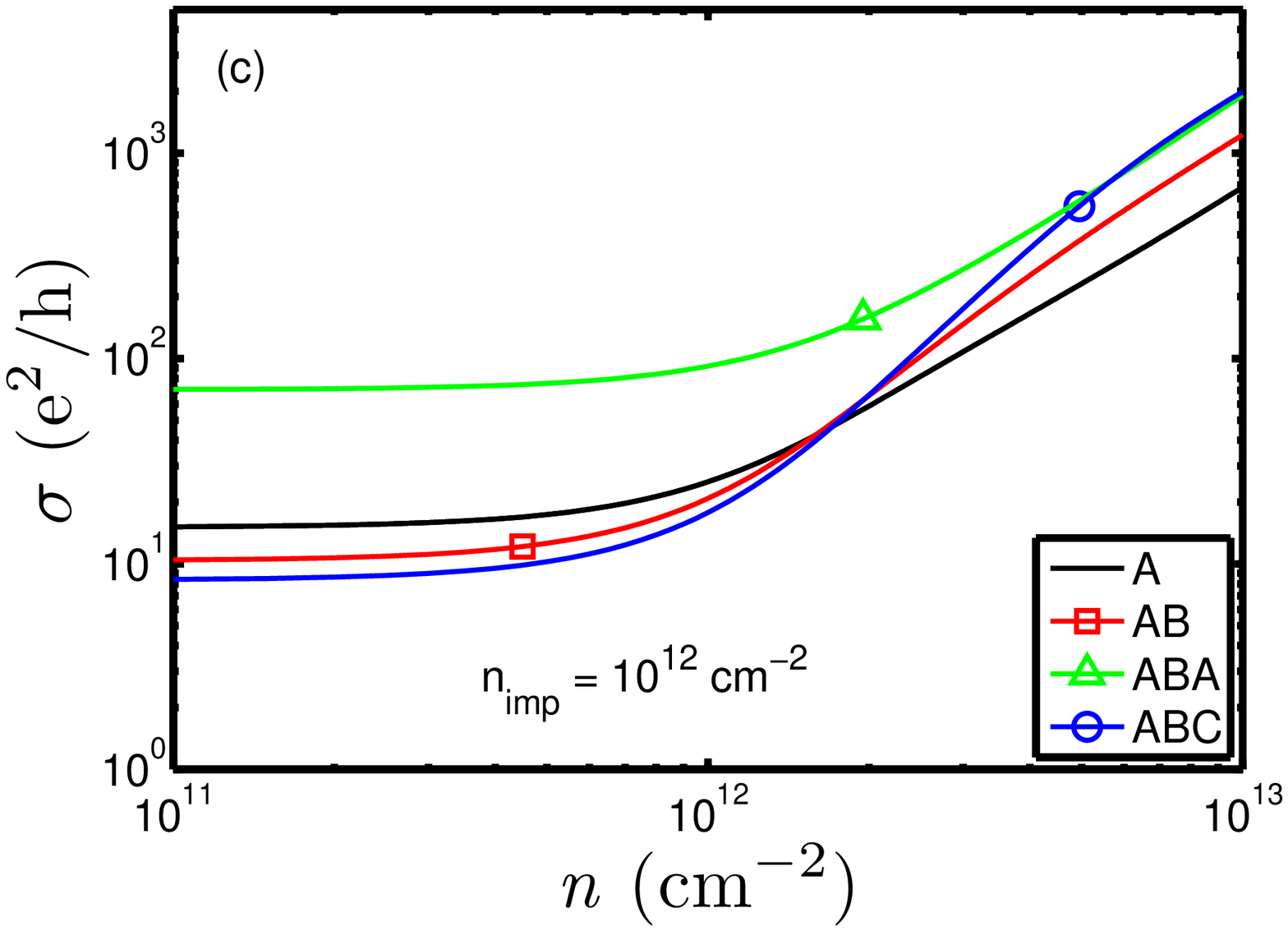} \\
\includegraphics[width=0.45\linewidth]{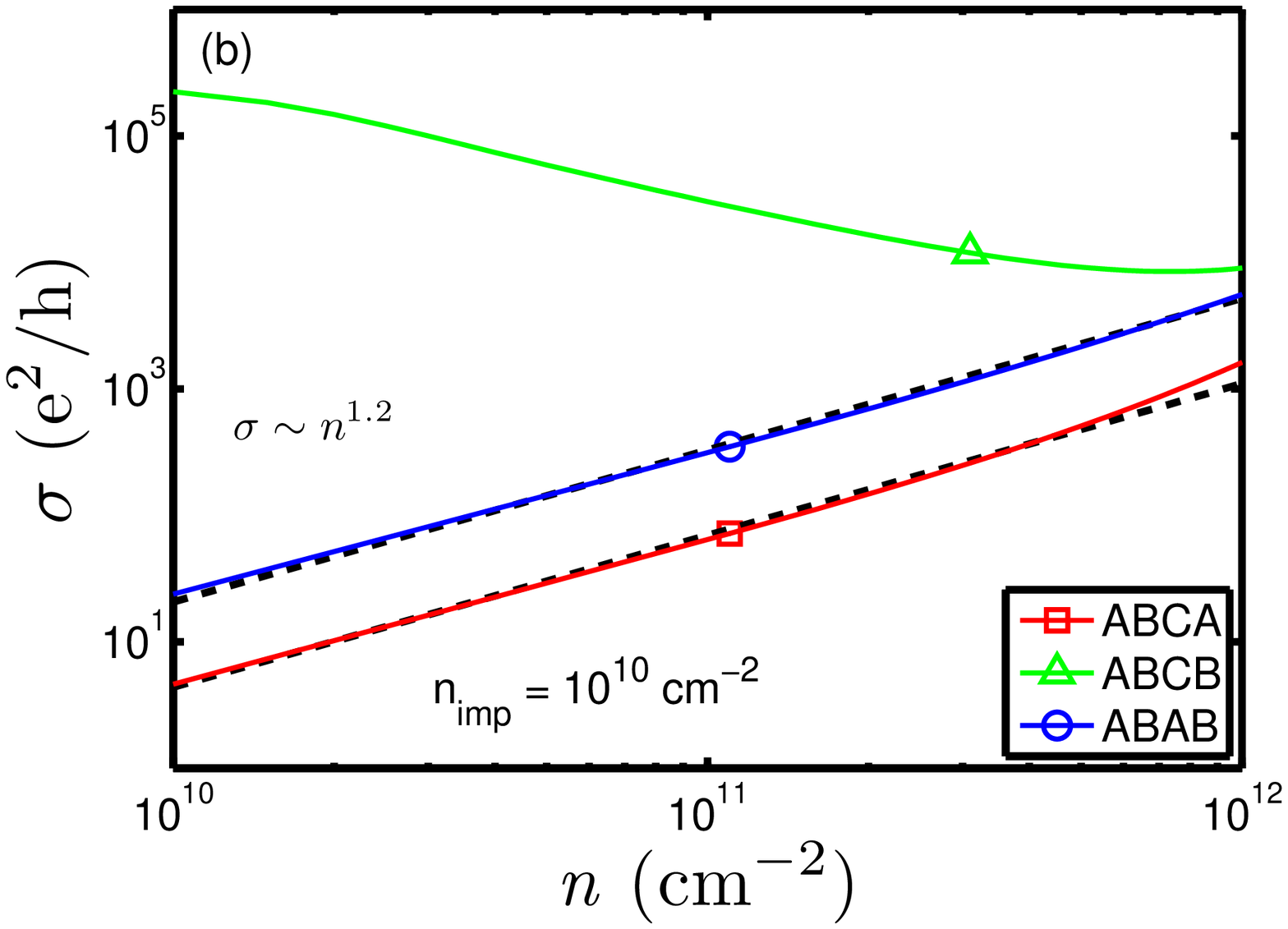} &
\includegraphics[width=0.45\linewidth]{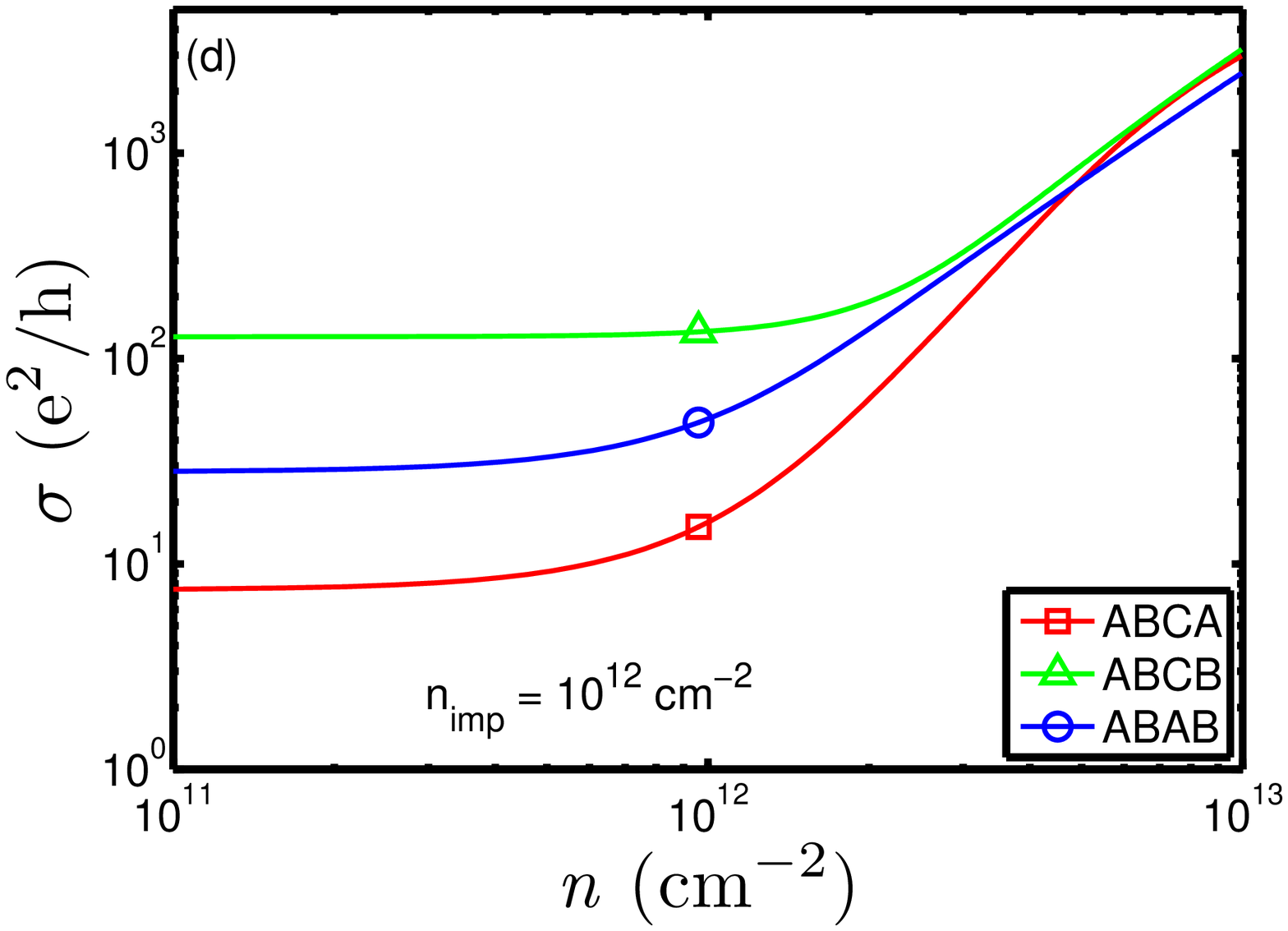}
\end{tabular}
\caption{\label{fig:long-range} (Color online) 
Graphene conductivity (solid lines) assuming charged impurity scattering.  
(a) For clean samples, the long-range Coulomb scatterers give similar power-law
dependence (dashed lines) of the conductivity on carrier density for monolayer,
bilayer and ABC trilayer graphene.  (b) The ABCA 
and ABAB  stacking orders of
tetralayer graphene give a similar power-law dependence 
(dashed lines) to monolayer and bilayer graphene.  We conclude that the low density transport 
properties for many graphene multilayers look quite similar under long-range scattering,
making it difficult to distinguish between them in a transport measurement.  (c) For dirty samples,
the transfer curves for monolayer, bilayer and trilayer graphene are
each quite different and can be easily distinguished.  (Notice the
logarithmic scale on the $y$-axis).  (d) For dirty tetralayer graphene,
the conductivity depends strongly on the stacking sequence, and
transport measurements could distinguish between the various types of
stacking.}
\end{center}
\end{figure*}

The results for charged impurities is shown in
Fig.~\ref{fig:long-range}.  Again the left panel shows a clean sample
($n_{\rm imp} = 10^{10}~{\rm cm}^{-2}$).  While all the different
stacking sequences have a power-law dependence for $\sigma(n)$,
unfortunately, with the exception of ABA and ABCB, the rest
are all very close to a linear dependence.  It
would therefore be difficult to distinguish the samples in any
transport measurement in the low carrier density regime (i.e. $n <
10^{12}~{\rm cm}^{-2}$).  For a dirty sample with a larger range of
carrier density ($n_{\rm imp} = 10^{12}~{\rm cm}^{-2},n < 10^{13}~{\rm
  cm}^{-2}$), we find that the functional forms for $\sigma(n)$ are
sufficiently different.  We emphasize that the figures show the
conductivity on a log-log scale.  Different slopes correspond to 
different scaling exponents $\gamma$, where $\sigma(n) \sim n^{\gamma}$.
For example, as  seen in Fig.~\ref{fig:long-range}d, while the 
different stackings of the tetralayer have similar values for 
the conductivity at $n=10^{13}~{\rm cm}^{-2}$, the minimum conductivity
for the ABCB tetralayer is two orders of magnitude larger than the ABCA stacking.  

To further understand these results, we note that the transport properties of 
graphene multilayers are
determined by two characteristic densities: the band density $n_0\equiv g_{\rm s} g_{\rm v} k_0^2/(4\pi)$ where $\hbar v_0 k_0=t_{\perp}/2$, 
and the impurity density $n_{\rm imp}$.  By gating graphene,
one can change both the carrier density and the type of carriers,
where a negative back-gate voltage induces holes, and a positive
back-gate induces electrons.  With special dielectrics one can induce
carrier densities as large as $n \approx 10^{14}~{\rm cm}^{-2}$ (see
Ref.~\onlinecite{kn:efetov2010}), although typically carrier densities
do not exceed $n \approx 5 \times 10^{12}~{\rm cm}^{-2}$.

For carrier densities much lower than the characteristic band density 
$n_0 \approx 2 \times 10^{12}~{\rm cm}^{-2}$, 
one can decompose the electronic structure of an arbitrary multilayer into parallel
channels of bands, each with the simple dispersion relation $\epsilon_k
\sim k^J$ where $J$ is the chirality index.  
The number of channels and the chirality index are
determined from the stacking sequence as discussed in
Ref.~\onlinecite{kn:min2008c}.  More details on the wave-functions and 
transport of $J$-chiral fermions can be found in Appendix~\ref{Ap:chirality}.

In the opposite limit of $n \gg n_0$, the band structure of $N$-layer
graphene decomposes into that of $N$ decoupled monolayer graphene
sheets, irrespective of the stacking sequence.  Since the transport of
monolayer graphene has been well studied (see
Ref.~\onlinecite{kn:dassarma2010a}), we do not explore this limit in any
detail.  Although we note, that even at the extremely large carrier
density $n \approx 10^{14}~{\rm cm}^{-2}$, one is not yet in the limit
of essentially decoupled sheets.

The second important scale is that of disorder.  Typical values of
$n_{\rm imp}$ vary from $10^{10}~{\rm cm}^{-2}$ (in suspended graphene)
to $5 \times 10^{12}~{\rm cm}^{-2}$.  Only when the carrier density is
much larger than $n_{\rm imp}$ can one use the usual semi-classical
Boltzmann transport theory.  When $n \lesssim n_{\rm imp}$, the
inhomogeneous landscape of electron and hole puddles gives rise to a
saturation in the conductivity approaching a finite $\sigma_{\rm min}$
at the Dirac point.

One can immediately identify two regimes that are experimentally
relevant.  When $n_{\rm imp} \ll n \ll n_0$, one can exploit the chiral
decomposition to obtain a power-law dependence of the conductivity on
carrier density.  This is what we called the ``clean sample'' regime 
(the left panels of Fig.~\ref{fig:short-range} and
Fig.~\ref{fig:long-range}).  In Appendix~\ref{Ap:chirality}
we derive the Effective Medium conductivity for arbitrary
$J$-chiral fermions, and we can use this decomposition to understand 
our results.  

As an example, consider the clean ABC stacked trilayer.  As seen in
Fig.~\ref{fig:short-range}, $\sigma(n) \sim n^{2}$ for short-range
impurities.  This dependence follows directly from the low energy
chiral decomposition discussed in Appendix~\ref{Ap:chirality} where
the ABC trilayer is approximated by a $J=3$ chiral system, and
$\sigma_J \sim n^{J-1}$ (see Eq.~\ref{sigmaJ}).  Similarly, the
numerical results for the ABC stacked trilayer with charged impurities
shown in Fig.~\ref{fig:long-range} is quite close to the expected
power-law $\sigma(n) \sim n$.  This small difference between the
numerical results and those anticipated from the chiral decomposition
is the result of our using a finite $d_{\rm imp} = 1~{\rm nm}$ for the
distance of the Coulomb impurities from the graphene sheet.  The
finite $d_{\rm imp}$ softens the small-distance divergence of the
Coulomb potential, thereby increasing the conductivity and giving a
larger coefficient $\gamma$ for the (approximate) power-law $\sigma
\sim n^{\gamma}$.

A similar analysis can be done for the ABA stacked trilayer graphene. At low energies, 
it is described by a direct product of $J=1$ and $J=2$ chiral systems.  At low 
density, the band structure has two sheets, one with a linear dispersion 
like monolayer graphene, and one with a parabolic dispersion  similar to 
bilayer graphene.  Requiring a constant Fermi energy, one can introduce
a dimensionless parameter $x= \varepsilon_{\rm F}/t_{\perp}$.  The chiral decomposition 
is valid when $x \ll 1$.  One notes that $x \approx (v^*/v_0)^J (k_{\rm F}/k_0)^J$ 
where the effective Fermi velocity $v^* \approx v_0$ is shown in Table~\ref{table}.
This implies that at low carrier density where the chiral decomposition is valid, the band with 
the larger chirality also has a larger carrier density and larger density of 
states.  In the case of ABA trilayers, this implies that for short-range 
scatterers, it behaves exactly like the AB bilayer which is also a $J=2$ chiral
system and $\sigma_{J=2} \sim n$ (as seen in Fig.~\ref{fig:short-range}a).  For 
charged impurities, the large density of states from the $J=2$ band effectively 
screens the impurities so that the $J=1$ band behaves like monolayer 
graphene with short-range impurities having $\sigma\sim$ constant.  All the 
shown power laws in left panels of Fig.~\ref{fig:short-range} and Fig.~\ref{fig:long-range} 
can be understood in this manner using the chiral decomposition.

The second regime relevant to experiments is when $n_{\rm imp}
\lesssim n_0 \ll {\rm max}(n)$.  We called this the ``dirty sample''
regime, since having a cleaner sample offers no qualitative
difference.  The important point is that since $n\gtrsim n_0$, the
chiral decomposition is not valid, and the band-structure has no
simple analytical form.  This regime can be seen in the right panels
of Fig.~\ref{fig:short-range} and Fig.~\ref{fig:long-range}.  Although
no simple analytical expression or power-law behavior determines
$\sigma(n)$, the results for the different number of layers, the different
stacking orders and short-range vs. long-range impurities are all
sufficiently different.  Therefore, by comparison of experimental data
to the results presented here, it should be possible to identify not only the
number of layers and stacking sequence, but also the nature of the
dominant source of disorder.

\section{Conclusion}
\label{conclude}

We have considered the transport properties of multilayer graphene
stacks.  The formalism is quite general and can be used for $N$-layers
of graphene with arbitrary stacking between the layers.  In the
absence of any experimental data for layers with $N>2$, we have
considered the most energetically favorable stacking sequences and the
cases of both short-range and long-range impurities.  For monolayer
and bilayer graphene, our results agree with previously known results
(see Appendix~\ref{Ap:bilayer}).  For trilayer graphene, we show that
ABA and ABC stacking have very different transport properties and
can be distinguished from each other.  Similarly, for tetralayer graphene, ABCA, ABCB
and ABAB each is a separate electronic material with its own
characteristic dependence of conductivity on carrier density.  (The
ABCB and ABAC stackings have the same conductivity since they are
related to each other by a symmetry transformation.)

An important objective of this work is to enable experimentalists working on
multilayer graphene to be able to use transport measurements to
identify and characterize their multilayer samples.  In addition, one
could use our results to determine the nature of the dominant impurity
potential,\cite{kn:xiao2010} the effect of changing the dielectric
environment~\cite{kn:jang2008} or identify when other effects (such as
quantum interference\cite{kn:adam2008d}) that we have neglected in our 
semi-classical approach become important.
  
Information from transport measurements could be used in conjunction with other
techniques such as Raman spectroscopy~\cite{kn:ferrari2006}, optical
detection~\cite{kn:min2009} and observing the scattering from phonons at high
temperature.~\cite{kn:min2010} Our main finding is that in most cases, the different
stacking sequences have different electronic properties that result in
characteristic dependence of the conductivity on the applied carrier
density.

\acknowledgements
The authors thank E. H. Hwang for valuable comments.
This work was supported in part by the NIST-CNST/UMD-NanoCenter 
Cooperative Agreement. 

\appendix

\section{Validity of the Born approximation}
\label{Ap:born}
In this work, we use the Born approximation for both short-range and Coulomb impurities as seen 
explicitly in Eq.~\ref{eq:squared_potential_short}.  This
approximation is valid if the scattering potential is weak enough.  As
the scattering potential becomes stronger, the scattering approaches
the unitary limit, in which the scattering amplitude becomes
independent of the strength of the potential.
In the unitary limit, the strength of the impurity potential drops
out of the expression for the conductivity, which then depends only on the impurity concentration~\cite{kn:ferreira2010}
\begin{equation}
\sigma =\frac{4 e^2}{h} \frac{k_{\rm F}^2}{2\pi^2 n_{\rm imp}} \ln^2(k_{\rm F} R).
\end{equation}
where $R$ characterizes the range of the potential.~\cite{kn:stauber2007}  In the opposite limit of 
weak scatterers, the strength of the impurity potential determines the
conductivity.  For consistency 
with the notation in Ref.~\onlinecite{kn:ferreira2010}, we introduce a scattering potential 
$V_{\rm imp}=V_0$ in Eq.~\ref{eq:tau} so that Eq.~\ref{eq:conductivity} reads
\begin{equation}
\sigma = \frac{g_{\rm s} g_{\rm v} }{2} \frac{e^2}{h} \frac{1}{n_{\rm imp}} \left( \frac{\hbar v_{\rm F}}{V_0} \right)^2.
\end{equation}
\noindent In the Born limit, it is the product $n_{\rm imp} V_0^2$ that determines the conductivity.  

Ultimately, microscopic models of measured
defects will determine whether the scattering is better described by
the unitary limit or the Born limit.
In the absence of such a model, an important issue for our results is
whether the combination of conductivities and carrier densities we
consider require that the scattering potential be so strong that the
Born approximation is no longer valid.
In a recent article, Ferreira {\it et al.}\cite{kn:ferreira2010} test
the validity of the Born approximation  
by computing the scattering amplitude as a function of the scattering
potential strength for short-range impurity scattering in monolayer and bilayer
graphene.  For weak scattering potentials, the Born approximation
results agree with their more general calculations but for large
enough potentials, the scattering amplitude reaches the unitary limit
where the scattering phase-shift is $\pi/2$ irrespective
of the strength of the impurity potential.
We use their results to argue that the
conductivities and charge densities we treat are consistent with the
Born approximation being valid.  

As discussed by Ferreira {\it et al}. the validity of the Born
approximation depends on the quantity $A = 
(V_0/\pi \hbar v_{\rm F}) k_{\rm F} \ln(k_{\rm F} R)$.  If $A \gg 1$,
we have the unitary limit, and if $A \ll 1$ we are in the Born limit.  To check 
the self-consistency of the Born approximation limit, we can re-write this 
condition as
\begin{eqnarray}
A^2 &=& \left[ \frac{V_0}{\pi \hbar v_{\rm F}} k_{\rm F} \ln(k_{\rm F} R)\right]^2 \approx 
\left[ \frac{V_0}{\pi \hbar v_{\rm F}} k_{\rm F} \right]^2 \nonumber \\
&\approx& 2 \frac{4}{\pi} \frac{n}{n_{\rm imp}} \frac{e^2/h}{\sigma} \ll 1,
\end{eqnarray}
where we assume $\ln(k_{\rm F} R) \approx 1$.  
A similar expression is obtained for bilayer graphene with the
prefactor $8/\pi$ replaced by $1/(4 \pi)$.  The Born approximation is
valid for a relatively high density of relatively weak scatterers.
Since $\sigma > 4e^2/h$
(both experimentally, and for the validity of the diffusive
approximation, see Ref.~\onlinecite{kn:adam2008d}), and typical
carrier densities range from $10^{10}~{\rm cm}^{-2}$ to $5\times
10^{12}~{\rm cm}^{-2}$, the Born approximation provides a consistent
solution when there are
no fewer than one short-range impurity per $40~{\rm nm}^2$ (or more
than one defect per 2000 carbon atoms).  These numbers seem quite
reasonable, given the preparation method of these samples.  
In Ref.~\onlinecite{kn:jang2008} the authors
measured $\sigma = 280~e^2/h$ for monolayer graphene, giving two
orders of magnitude wider 
range of carrier densities for the validity of the Born approximation
for the same impurity concentration.  Similarly, in Fig.~\ref{fig:short-range}a 
although we take $n_{\rm imp} = 10^{10}~{\rm cm}^{-2}$ in order to illustrate the
power law dependence of the conductivity due to the chiral
decomposition, the Born approximation is still valid because we also
take $\sigma/(e^2/h)$ to be much larger.  While such arguments
do not rule out the possibility that the scatterers are in the  unitary
limit, they nonetheless 
establish the Born approximation treatment is self-consistent.

\section{Conductivity of $J$-chiral Fermions}
\label{Ap:chirality}

At very low carrier density, an arbitrarily stacked graphene multilayer
can be described as a superposition of pseudospin doublets.  This
decomposition holds so long as 
$\hbar v_0 k_{\rm F} \ll t_{\perp}$, where $J$ is the chirality index for the pseudospin doublet.  

The rules for the decomposition are as follows: (monolayer graphene)
A$\rightarrow (J=1)$; (bilayer graphene) AB$\rightarrow (J=2)$;
(trilayer graphene) ABA$\rightarrow (J=1) \oplus (J=2)$ and ABC
$\rightarrow (J=3)$.  This notation indicates, for example, that an
ABC stacked trilayer is described by a $3$-chiral Hamiltonian, while
an ABA stacked trilayer is composed of two bands -- one similar to
monolayer graphene and the second similar to bilayer graphene (see
Table~\ref{table} and Refs.~\onlinecite{kn:min2008c,kn:min2008d,kn:koshino2010} 
for more
details).  The $J$-chiral Hamiltonian is defined as
\begin{equation}
{\cal H}=t_{\perp}\left(
\begin{array}{cc}
0 &(\nu_{\bm{k}}^{\dagger})^J \\
(\nu_{\bm{k}})^J &0 \\
\end{array}
\right),
\label{H_J}
\end{equation}
where $\nu_{\bm{k}}\equiv\hbar v^{\ast} k e^{i\phi_{\bm{k}}} /t_{\perp}$ and $v^{\ast}$ is the effective in-plane Fermi velocity (for example, $v^{\ast}=v_0$ for $J=1$ monolayer graphene and $J=2$ bilayer graphene, and in general for periodic ABC stacking).  

Throughout the manuscript we have used the following properties of the
energy levels and wavefunctions for the $J$-chiral system
\begin{eqnarray}
\varepsilon_{s,k}&=&s t_{\perp} \left(\hbar v^{\ast} k \over t_{\perp}\right)^J, \\
\left|s,\bm{k}\right>&=&
{1 \over \sqrt{2}}\left(
\begin{array}{c}
s \\
e^{i J\phi_{\bm{k}}}\\
\end{array}
\right).\nonumber
\label{Ek_J}
\end{eqnarray}
\noindent
The band-index $s=\pm 1$ corresponds to the positive (negative) energy states 
of the conduction (valence) band and $\varepsilon_{\rm F} = \varepsilon_{s,k=k_{\rm F}}$.  
The intraband chirality factor is then calculated as 
\begin{equation}
F(\phi)=|\left<s,k,\phi=0|s,k,\phi\right>|^2={1\over 2}\left[1+\cos (J\phi)\right].
\label{Fphi_J}
\end{equation}

The scaling of the conductivity with carrier density can be immediately
obtained by noticing that $v_{\rm F}\sim k_{\rm F}^{J-1}$ and
$\rho(\epsilon_F)\sim k_{\rm F}^{2-J}$.  This gives
\begin{equation}
\sigma_J \sim {n^{J-1} \over n_{\rm imp} V_{\rm imp}^2},
\end{equation}
which depends on the scattering potential $V_{\rm imp}$.  
Assuming $d_{\rm imp}=0$ for simplicity and restoring the
dimensions, we find
\begin{widetext}

\begin{table}[t]
\caption{$J$-chiral decomposition for monolayer, bilayer, trilayer
and tetralayer graphene with different stacking arrangements 
(see Ref.~\protect{\onlinecite{kn:min2008c}} for more details).}
\begin{center}

\begin{tabular}{c|c|c|c}
\hline\hline
\ Number of layers (N) \ & \ Stacking sequence \ & \ Chirality (J) \ & \ Effective velocity $v^*/v_0$\ \\
\hline
1 & A    & 1          & $1$\\
2 & AB   & 2          & $1$\\
3 & ABA  & 1$\oplus$2 & $1 \oplus 2^{-1/4}$\\
3 & ABC  & 3          & $1$\\
4 & ABAB & 2$\oplus$2 & $1/\sqrt{\left(\sqrt{5}+1\right)/2} \oplus 1/\sqrt{\left(\sqrt{5}-1\right)/2}$\\
4 & ABCA & 4          & $1$\\
4 & ABCB & 1$\oplus$3 & $1 \oplus \sqrt{2}/2 $\\
4 & ABAC & 1$\oplus$3 & $1 \oplus \sqrt{2}/2 $\\
\hline   
\end{tabular}
\label{table}
\end{center}
\end{table}

\begin{eqnarray}
\sigma_J(n) = \left\{
\begin{array}{ll}
{e^2\over h} \left({n\over n_{\rm imp}}\right) \left[\left({\hbar v^{\ast} \over t_{\perp}}\right)^2  {4 \pi n \over g_{\rm s} g_{\rm v}}\right]^{J-2} \left({\hbar v^{\ast} \over t_{\perp} d_{\rm sc}}\right)^2 {J^2  \over \pi\alpha^2 \beta_J} \propto n^{J-1}, & \ \ \mbox{Short-range disorder}, \\
{e^2\over h} \left({n\over n_{\rm imp}}\right) \left[\left({\hbar v^{\ast} \over t_{\perp}}\right)^2 {4 \pi n \over g_{\rm s} g_{\rm v}} \right]^{J-1} {2 J^2 \over \pi\alpha^2 \gamma_J} \propto n^{J}, & \ \ \mbox{Bare Coulomb}, \\
{e^2\over h} \left({n\over n_{\rm imp}}\right) {16 \over \pi \beta_J}\propto n, & \ \ \mbox{Overscreened Coulomb ($\alpha\gg 1$)},
\end{array} \right.
\label{sigmaJ}
\end{eqnarray}
\end{widetext}
\noindent where $\beta_J=1/2$ for $J=1$, $\beta_J=1$ for $J>1$ and $\beta_J=2$ for $F(\phi)=1$,
while $\gamma_J=1$ for $F(\phi)$ in Eq.~(\ref{Fphi_J}) and $\gamma_J=2$ for $F(\phi)=1$. (Here we are considering a $J$-chiral system in Eq.~\ref{H_J} and did not include intervalley scatterings.)
The result for the bare Coulomb potential was presented in Eq.~\ref{sigmaJ} for a pedagogical reasons, and $F(\phi)=1$ case for the bare and screened Coulomb potentials was also considered for completeness. 
We note that since 
\begin{equation}
\tilde{q}_{\rm TF}\equiv {q_{\rm TF} \over k_{\rm F}}={4\alpha \over J} \left({\hbar v^{\ast} k_{\rm F} \over t_{\perp}}\right)^{1-J},
\end{equation}
for the low density limit $k_{\rm F} \rightarrow 0$, the overscreened
Coulomb potential becomes a good approximation for charged impurities.
In this low density limit, among the short-range scattering and screened Coulomb scattering, 
screened Coulomb scattering dominates over
short-range scattering for $J < 2$, (with a corresponding
$\sigma(n) \sim n$); while short-range scatterers dominate for $J
> 2$ (and $\sigma(n) \sim n^{J-1}$). 
(Note that because of the Matthiessen's rule, a scattering mechanism with smaller conductivity dominates.)
Bilayer graphene at low
carrier density (or $J=2$) is unique in that both charged impurities
and short-range disorder give conductivities with the same carrier
density dependence~\cite{kn:adam2008a,kn:katsnelson2007b,kn:dassarma2010} 
making them difficult to
distinguish experimentally.\cite{kn:xiao2010}  In the opposite limit
of very high carrier density, the energy band structure of multilayer
graphene separates into a collection of decoupled monolayer graphene
bands.\cite{kn:min2009} As a result, the conductivity scales as that
of a monolayer at very high carrier density.

As discussed above, the chiral decomposition works only at low carrier
density where it is known that density fluctuations dominate the 
transport properties.\cite{kn:hwang2006c,kn:adam2007a,kn:rossi2008b}
One must therefore estimate whether there is a regime of validity where the 
carrier density is large enough so that the puddle physics no 
longer dominates (i.e. $n \gg n_{\rm rms}$), but small enough that
the chiral decomposition is still valid (i.e. $n \ll n_0$); here
$n_{\rm rms}$ is the root-mean-square fluctuation in carrier density induced
by the disorder potential, while 
$n_0 \approx 2~\times~10^{12}~{\rm cm}^{-2}$ is the crossover density scale.  
We estimate $n_{\rm rms}$ using the self-consistent approximation of 
Refs.~\onlinecite{kn:adam2007a} and \onlinecite{kn:adam2008a},
where $\langle \varepsilon_{\rm F}^2 \rangle = n_{\rm imp} \langle V_{\rm D}^2 \rangle$,
and $V_{\rm D}$ is the disorder potential of screened charged impurities located
at some distance $d_{\rm imp}$ from the graphene sheet
\begin{eqnarray}
\langle \varepsilon_{\rm F}^2 \rangle &=& n_{\rm imp} \int \frac{d^2q}{(2 \pi)^2} 
\left[ \frac{2 \pi e^2 \exp(-q d_{\rm imp})}{\epsilon (q + q_{\rm TF})} \right]^2,
\nonumber \\
&=& 2 \pi n_{\rm imp} \alpha^2 (\hbar v_0)^2 C_0^{\rm TF}(2 q_{\rm TF} d_{\rm imp}).
\end{eqnarray}
\noindent For the relevant limit $n \ll n_0$, $C_0^{\rm TF}(x) \approx x^{-2}$ 
(for details, see Ref.~\onlinecite{kn:adam2007a}).  Although we used a
charged impurity model for the disorder potential, in this limit the
impurities are perfectly screened, and the long-range nature of the
impurity becomes irrelevant (i.e. one gets similar results starting
from a short-range impurity model).  
Assuming a Gaussian probability distribution for the carrier density,
we find that for the $J$-chiral
Hamiltonian (with $J\geq 2$)
\begin{equation}
\label{eq:nrms}
n_{\rm rms} \approx \sqrt{\frac{3 n_{\rm imp} J^2}{32 \pi d_{\rm imp}^2}},
\end{equation}
which shows that for any given $J$, one can determine the minimum disorder
concentration $n_{\rm imp}$ necessary to ensure that $n \gg n_{\rm rms}$.  
(The factor of 3 inside the square-root is added to conform to the convention for graphene monolayers, see Ref.~\onlinecite{kn:dassarma2010a}.)
Note that for $J=1$ monolayer graphene, one has to use the full dielectric function for the screening.\cite{kn:adam2007a}

\begin{figure}[!ht]
\bigskip
\begin{center}
\includegraphics[width=0.9\linewidth]{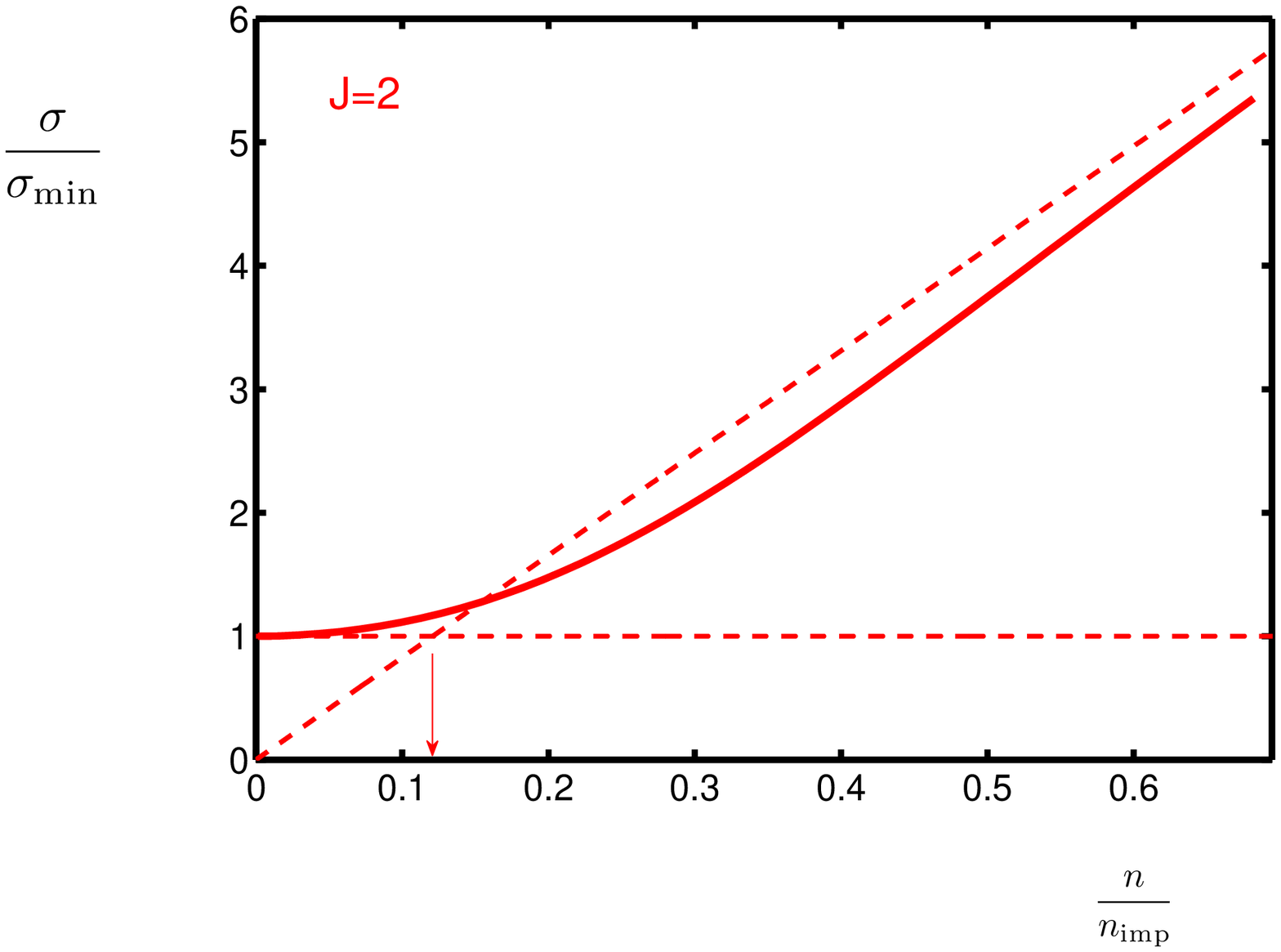}
\includegraphics[width=0.9\linewidth]{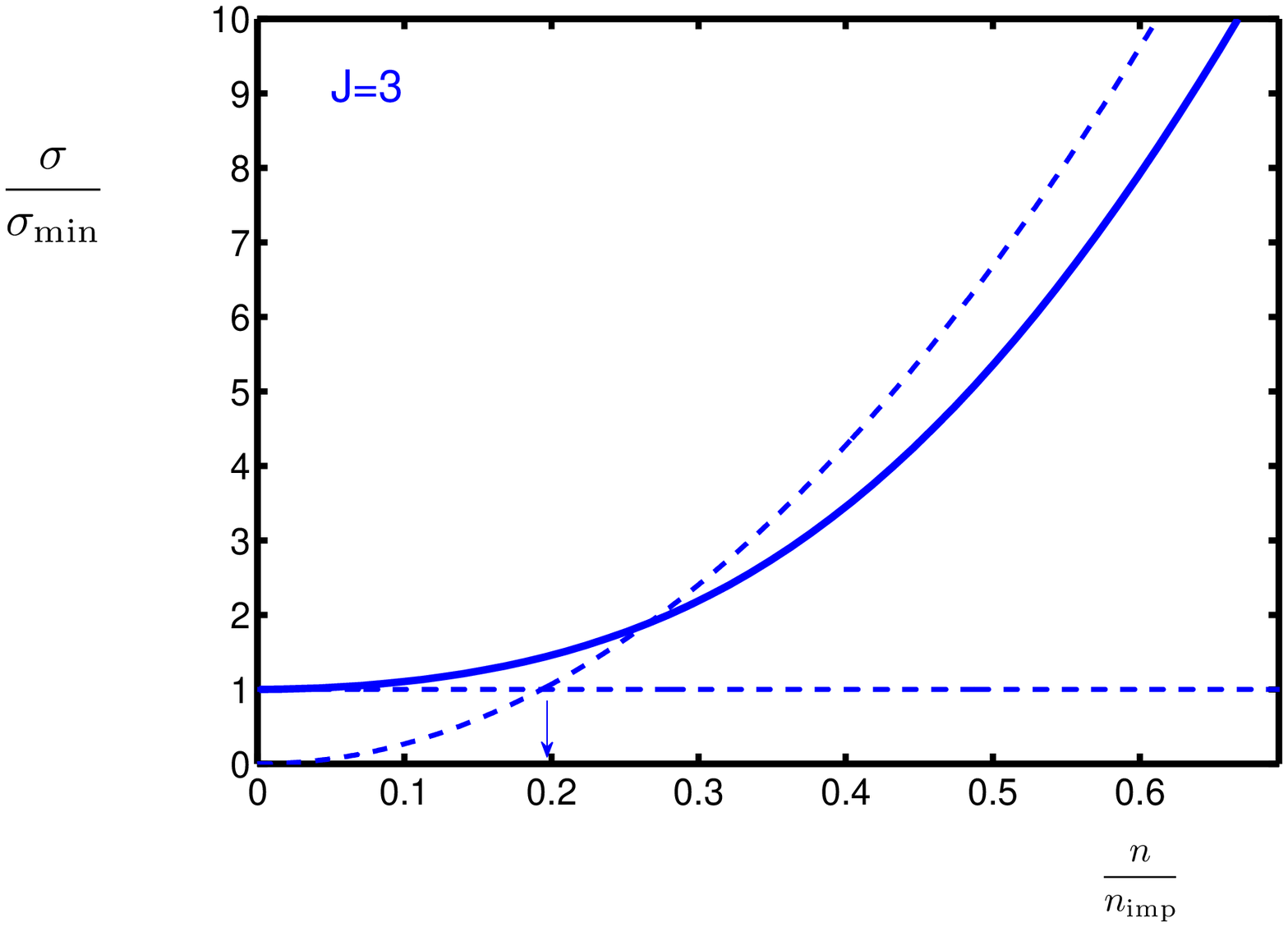}
\includegraphics[width=0.9\linewidth]{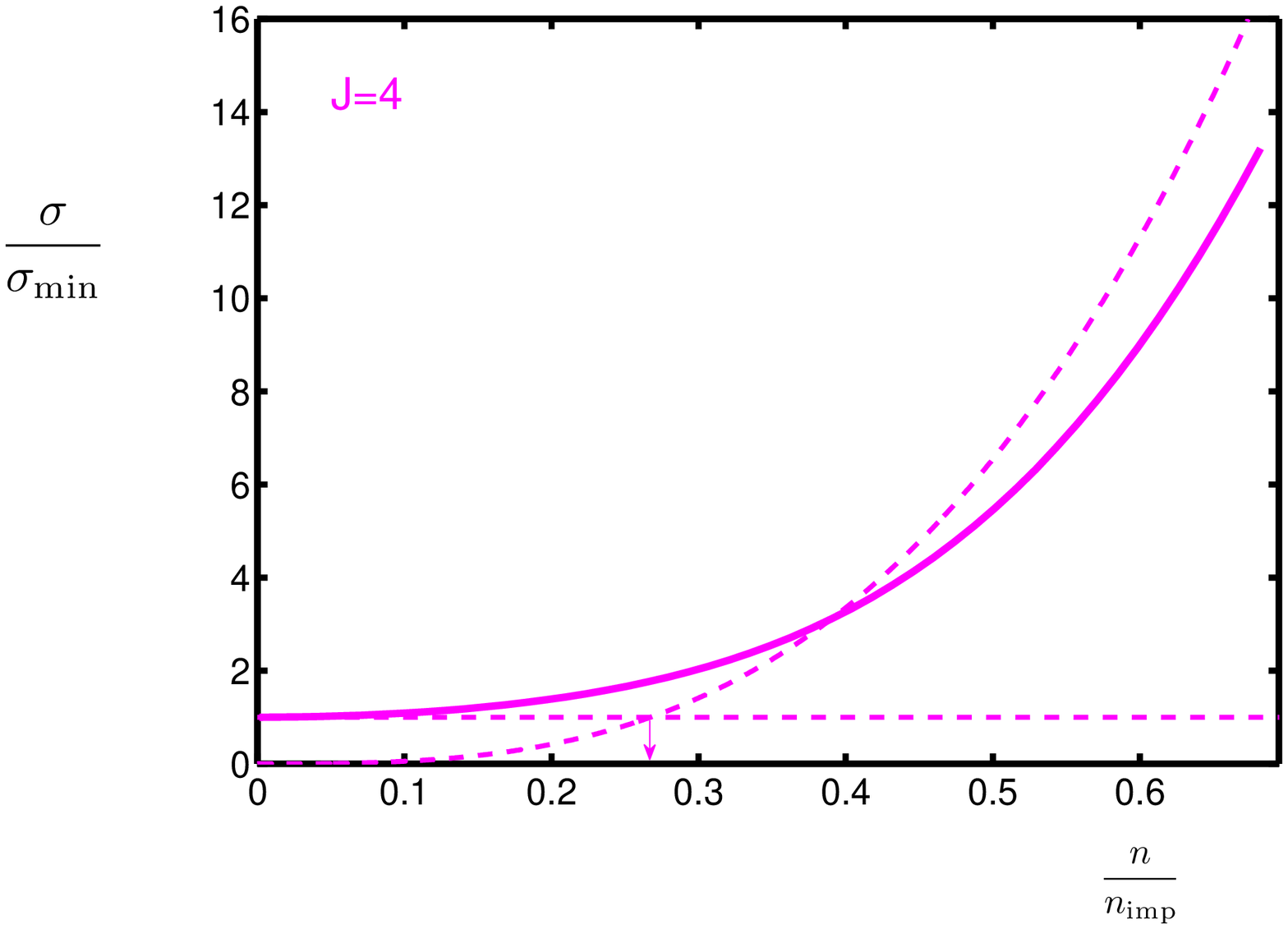}
\end{center} 
\caption{\label{fig:emt} (Color online) Effective medium theory
result (solid lines) for the conductivity 
assuming short-range disorder
as a function of 
dimensionless density $n/n_{\rm imp}$
for the $J$-chiral Hamiltonian (see Eq.~\ref{eq:emt}) for
$J=2,3$ and $4$.
The results assume an impurity density of $n_{\rm imp}=10^{11}$~cm$^{-2}$. 
Note that $\sigma/\sigma_{min}$ is independent of $d_{\rm sc}$ because $d_{\rm sc}$ appears as a multiplicative factor in Eq.~\ref{sigmaJ}.
Dashed lines show minimum conductivity
(i.e. $\sigma/\sigma_{\rm min} = 1$), and the Boltzmann conductivity (Eq.~\ref{sigmaJ})
without performing the EMT average.  Notice that the puddle regime (marked 
with arrows) gets larger
with increasing $J$.}
\end{figure}

From Eqs.~\ref{eq:emt1} and \ref{sigmaJ} one can easily construct an Effective 
Medium Theory for the $J$-chiral model. The conductivity 
is obtained by solving the integral equation
\begin{equation}
\int_0^{\infty} dn'~ \exp \left[ \frac{-n'^2}{2 n_{\rm rms}^2} \right]
\cosh \left[ \frac{n n'}{n_{\rm rms}^2} \right]
\frac{\sigma_J(n')-\sigma_{\ty{EMT}}(n)} {\sigma_J(n')+ \sigma_{\ty{EMT}}(n)} = 0.
\label{eq:emt}
\end{equation}

To illustrate the differences between different $J$-chiral models, in
Fig.~\ref{fig:emt} we show $\sigma_{\ty{EMT}}/{\rm min}(\sigma_{\ty{EMT}})$ for $J=2,3$
and $4$ 
assuming short-range disorder, 
where we estimate $n_{\rm rms}$ from Eq.~\ref{eq:nrms}
assuming that $d_{\rm imp} = 1~{\rm nm}$ and $n_{\rm imp} = 10^{11}~{\rm cm}^{-2}$.  As seen in the figure, the electron and hole puddles
tends to pin the conductivity value close to its minimum value, and
that the puddle regime increases with increasing $J$.

\section{Chirality factor for intervalley and interband scattering}
\label{Ap:chiralfactor}

\begin{figure}[!ht]
\bigskip
\begin{center}
\includegraphics[width=0.9\linewidth]{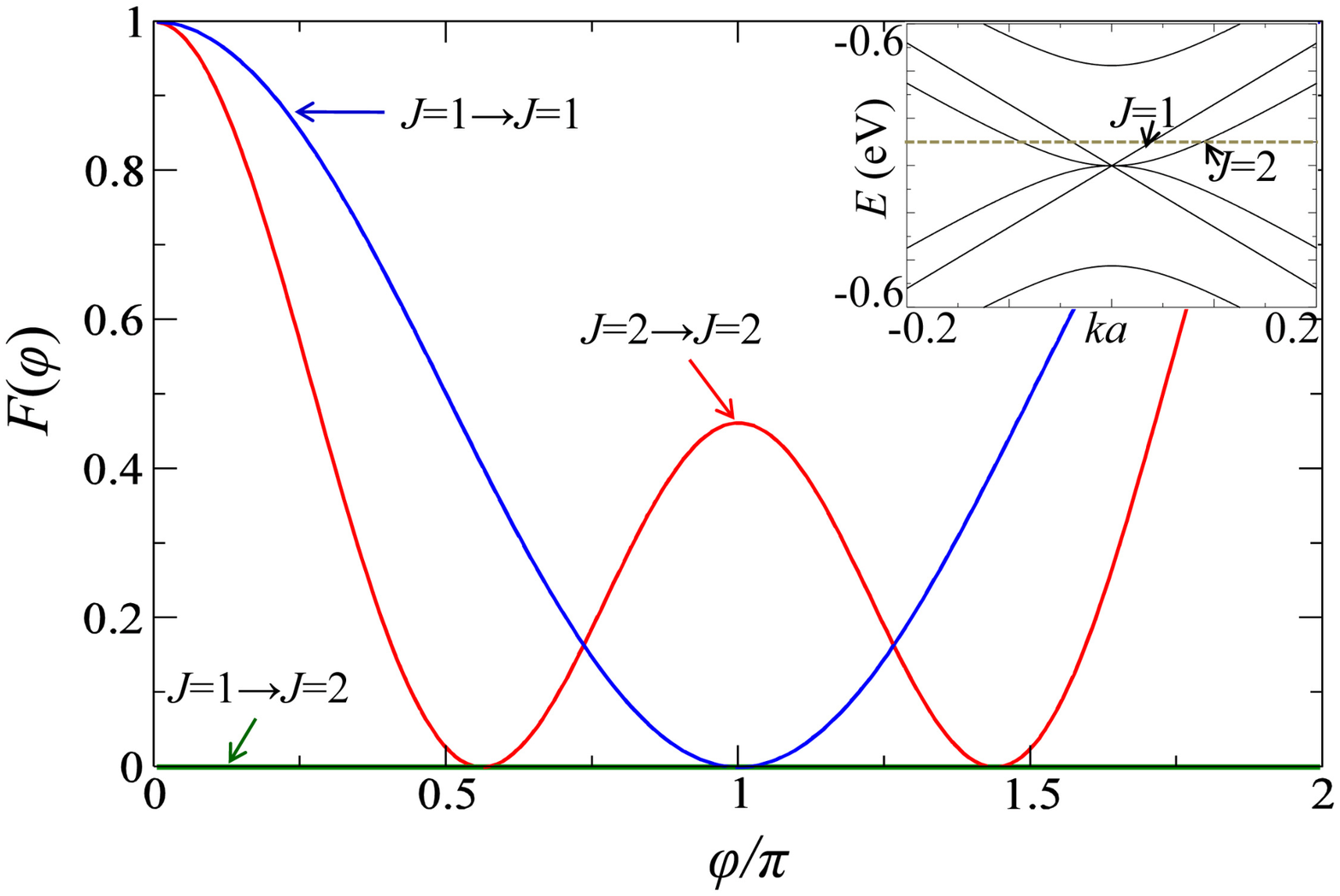}
\includegraphics[width=0.9\linewidth]{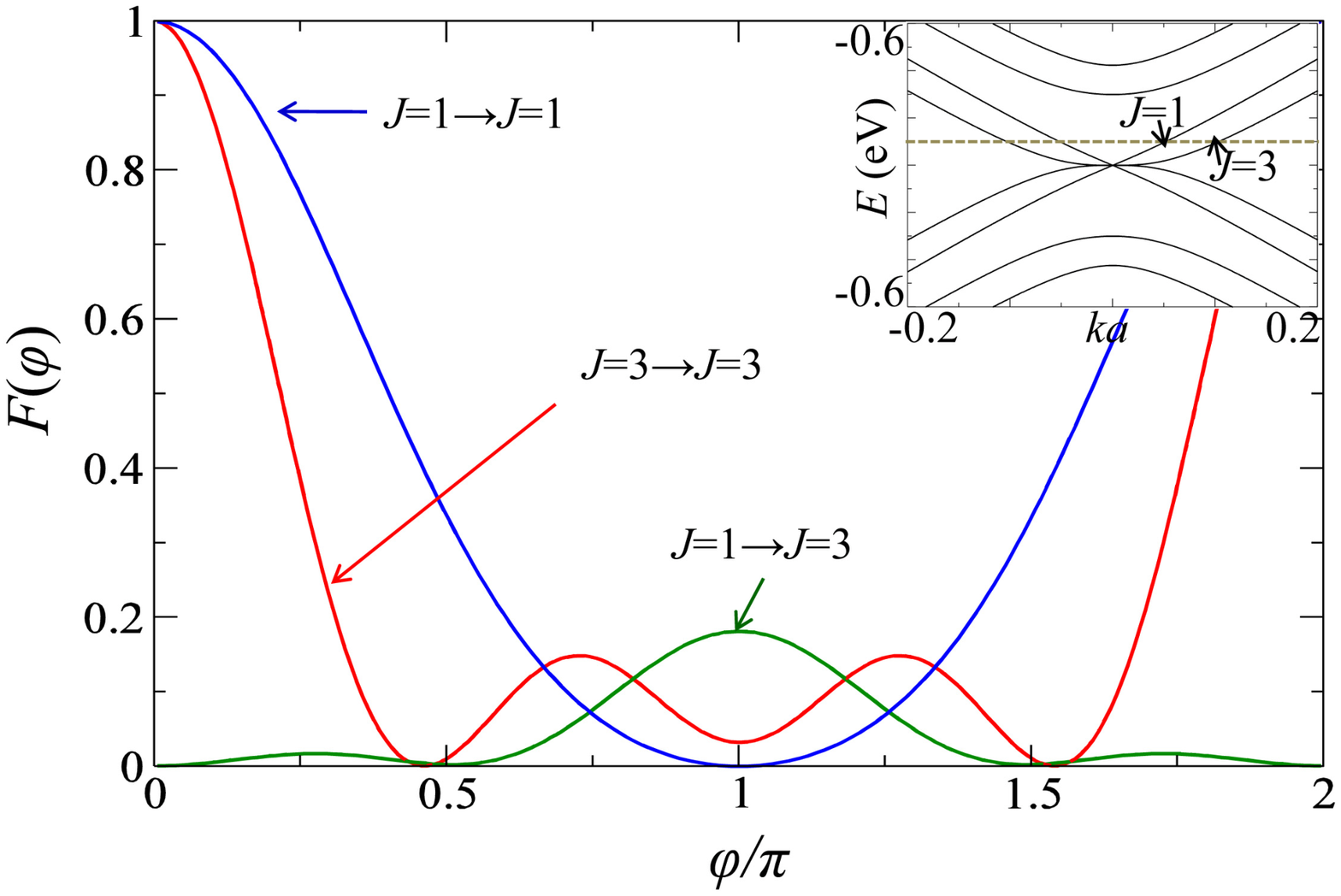}
\end{center} 
\caption{\label{fig:Fphi} (Color online)
The chirality factor $F(\phi)=|\langle s, {\bf k}, \phi=0 | s, {\bf k'}, \phi \rangle|^2$ is 
determined by the wavefunction overlap between 
initial and final states.  Top panel shows ABA stacked 
trilayer graphene, and bottom panel shows ABCB stacked tetralayer graphene for
$E_{\rm F} = 0.1~{\rm eV}$.  The bands are labeled according to the $J$-chiral decomposition 
shown in Table~\protect{\ref{table}}. For intraband scattering, the chirality factor at small
Fermi energy is given by Eq.~\protect{\ref{Fphi_J}}.  However, the interband chirality
factor is identically zero for scattering between the $J=1$ and $J=2$ bands of the ABA trilayer 
and strongly suppressed for scattering between the $J=1$ and $J=3$ bands of the ABCB 
tetralayer.  As discussed in the text, we believe that this suppression is unphysical 
for generic short-range impurities.  The insets
in both panels show the tight-binding band structure determined numerically by 
solving Eq.~\protect{\ref{eq:Hamiltonian}} as discussed in the text and dashed lines
indicate the Fermi energy.}
\end{figure}

It is often argued that monolayer graphene has a high mobility because the
chiral nature of carriers prevents backscattering.  In the diffusive regime,
the scattering rate involves an integral of the chirality factor $F(\phi)$ 
over all angles weighted by the Boltzmann factor $1- \cos\phi$.  As discussed
in Ref.~\onlinecite{kn:adam2008a}, the enhancement due to chirality is no more 
than a factor of order unity.  Similarly for graphene multilayers, one can calculate 
$F(\phi)$ numerically, and illustrative examples are shown in Fig.~\ref{fig:Fphi}
for the ABA trilayer and ABCB tetralayer.  For intraband scattering and at low
density, the chirality factor agrees with the analytic results in Eq.~\ref{Fphi_J} 
derived for $J$-chiral Hamiltonians.  For the range of carrier densities we consider, 
the chirality factor changes the conductivity by a factor of order unity, 
just like the case of monolayer graphene.  However, a similar calculation of
the interband chirality factor shows quite different results.  Figure \ref{fig:Fphi} shows that the 
interband chirality factor is exactly zero for the ABA trilayer and strongly 
suppressed for the ABCB tetralayer.  

It can be shown that the interband chirality factor in all periodic AB stackings vanishes from the form of the wavefunctions \cite{kn:min2008d},  
and in all other layer stackings it is strongly suppressed compared with the intraband chiral factor.  
At first glance, this might suggest 
that interband scattering
is negligible and can be ignored.  However, we point out that being able to decompose the
scattering matrix element into a plane-wave-like overlap $V_{\rm imp}^2$ and a chirality factor
$F(\phi)$ as we did in Eq.~\ref{eq:squared_potential} relies on the impurity potential 
being diagonal in the $2N \times 2N$ chiral-basis for $N$ graphene layers with 2 valleys.  This 
might be a reasonable assumption for the potential of remote charged impurities located at 
$d_{\rm imp} = 1$ nm to 2 nm away, but not for vacancies or adsorbates that would be 
strongly localized on one of the layers.  Without a microscopic theory that would give the 
matrix structure of the impurity potential in this chiral basis (and we note that such 
a theory would likely be non-universal depending strongly on the type of defect), it is 
more reasonable to set $F(\phi) = 1$ for short-range impurities.  For generic short-ranged
disorder, this would correctly weight the relative importance of the interband and intraband 
contributions at the expense of losing the chirality enhancement factor of order unity.  Since
we are interested in quantities such as the power-law trends for $\sigma(n)$ (left panels 
of Fig.~\ref{fig:short-range} and Fig.~\ref{fig:long-range}) or the ratio between
$\sigma(n)$ at high and low carrier density (right panels of 
Fig.~\ref{fig:short-range} and Fig.~\ref{fig:long-range}), this approximation 
is well suited to the scope of this work.         

\section{Bilayer graphene: Analytical results}
\label{Ap:bilayer}
In this section we derive analytic results for the transport 
properties of bilayer graphene.  The bilayer graphene 
Hamiltonian is 
\begin{equation}
{\cal H}=\left(
\begin{array}{cccc}
0 &v_0\pi^{\ast} &0 &0 \\
v_0\pi &0 &t_{\perp} &0 \\
0 &t_{\perp} &0 &v_0\pi^{\ast}\\
0 &0 &v_0\pi &0\\
\end{array}
\right),
\end{equation}
where $\pi=\hbar(k_x + i k_y)$ and energy eigenvalues 
and eigenvectors (up to normalization) are
\begin{equation}
\varepsilon_h^{-}=-\varepsilon_h, \, \varepsilon_l^{-}=-\varepsilon_l, \, \varepsilon_l^{+}=+\varepsilon_l, \, \varepsilon_h^{+}=+\varepsilon_l,
\end{equation}
\begin{eqnarray}
\left|\phi_h^{-}\right>&=&
\left(
\begin{array}{c}
v_0\pi^{\ast}\\
-\varepsilon_h\\
+\varepsilon_h\\
-v_0\pi\\
\end{array}
\right),
\,
\left|\phi_l^{-}\right>=
\left(
\begin{array}{c}
v_0\pi^{\ast}\\
-\varepsilon_l\\
-\varepsilon_l\\
v_0\pi\\
\end{array}
\right), 
\\
\left|\phi_l^{+}\right>&=&
\left(
\begin{array}{c}
v_0\pi^{\ast}\\
+\varepsilon_l\\
-\varepsilon_l\\
-v_0\pi\\
\end{array}
\right),
\,
\left|\phi_h^{+}\right>=
\left(
\begin{array}{c}
v_0\pi^{\ast}\\
+\varepsilon_h\\
+\varepsilon_h\\
v_0\pi\\
\end{array}
\right), \nonumber
\end{eqnarray}
where 
\begin{eqnarray}
\varepsilon_l&=&-t_{\perp}/2+\sqrt{(t_{\perp}/2)^2+(\hbar v_0 k)^2}, \\
\varepsilon_h&=&+t_{\perp}/2+\sqrt{(t_{\perp}/2)^2+(\hbar v_0 k)^2}. \nonumber
\end{eqnarray}

From Eq.~\ref{eq:conductivity} taking into account only the low energy 
band with the energy $\varepsilon_l^{+}$, the Fermi velocity is given by
\begin{equation}
v_{\rm F}={1\over \hbar}\left.{d\varepsilon\over dk}\right|_{\varepsilon=\varepsilon_{\rm F}}=v_0 {\hbar v_0 k_{\rm F} \over \sqrt{(t_{\perp}/2)^2+(\hbar v_0 k_{\rm F})^2}},
\end{equation}
and the density of states per spin and valley at the Fermi energy is
\begin{equation}
\rho(\varepsilon_{\rm F})={k_{\rm F} \over 2\pi\hbar v_{\rm F}}={{\sqrt{(t_{\perp}/2)^2+(\hbar v_0 k_{\rm F})^2}} \over 2\pi(\hbar v_0)^2}.
\end{equation}
The chirality factor within the same band $\varepsilon_l^{+}$ is given by~\cite{kn:xiao2010}
\begin{eqnarray}
\label{eq:chiral_factor_bilayer}
F(\phi)&=&|\left<\varepsilon_l^{+},\phi=0|\varepsilon_l^{+},\phi\right>|^2 \nonumber \\
&=&{1\over 4}\left[1-\eta+(1+\eta)\cos \phi\right]^2, 
\end{eqnarray}
where $\eta=1/\sqrt{1+n/n_0}$, $n=g_{\rm s} g_{\rm v} k_{\rm F}^2/(4\pi)$, $n_0=g_{\rm s} g_{\rm v} k_0^2/(4\pi)$ and $\hbar v_0 k_0=t_{\perp}/2$. 

For simplicity, let's consider the conductivity when the Fermi energy crosses only the low energy band $\varepsilon_l^{+}$.
For short-range disorder, from Eq.~\ref{eq:squared_potential_short}
\begin{equation}
\tilde{V}_{\rm imp}^2=k_{\rm F}^2 d_{\rm sc}^2 f(\eta)
\end{equation}
where $f(\eta)={1\over 8}(5\eta^2-2\eta+1)$ with the chirality factor of Eq.~\ref{eq:chiral_factor_bilayer}, while $f(\eta)=1$ for $F(\phi)=1$.
Then, we find 
\begin{equation}
\sigma={e^2\over h} \left({n\over n_{\rm imp}}\right) \left({\hbar v_0 \over t_{\perp}d_{\rm sc}}\right)^2  \left({1 \over \pi\alpha^2}\right) \left({2\eta^2\over f(\eta)}\right).
\end{equation}
Note that if we include the intervalley scattering, the conductivity becomes smaller by a factor of 2.

\noindent Similarly for the bare Coulomb disorder with $d_{\rm imp}=0$ we have 
\begin{equation}
\sigma={e^2\over h} \left({n^2\over n_{\rm imp} n_0}\right) \left({1 \over \pi\alpha^2}\right) \left({8\eta^2\over 3\eta^2-2\eta+3}\right),
\end{equation}
while for screened Coulomb disorder with $d_{\rm imp}=0$ we find (here $\alpha \gg 1$)
\begin{equation}
\sigma\approx{e^2\over h} \left({n\over n_{\rm imp}}\right) \left({64 \over \pi}\right) \left({1\over 5\eta^2-2\eta+1}\right).
\end{equation}
These results are consistent with the numerical data shown 
in Fig.~\ref{fig:short-range} and Fig.~\ref{fig:long-range} as 
well as with earlier results.\cite{kn:xiao2010,kn:adam2009b}
   

\end{document}